    \documentclass[12pt,psf,epsf]{article}
\usepackage[centertags]{amsmath}
\usepackage{amssymb}
\usepackage{graphicx}
\usepackage{epsfig}
\usepackage{ulem}
\usepackage[english]{babel}
\usepackage{array}
\usepackage{amsthm}
\usepackage{latexsym}
\usepackage[mathcal]{euscript}
\pdfoutput=1
\usepackage{epsfig}

 \usepackage{jheppub}
 \usepackage{hyperref}

\usepackage{subfigure}
\usepackage{psfrag}

\usepackage{epsfig}
\usepackage[latin1]{inputenc}
\usepackage{float}
\usepackage{graphicx}
\usepackage{cancel}
\usepackage{mathrsfs}
\usepackage{amssymb}
\usepackage{amsfonts}
\usepackage{amsmath}
\usepackage{slashed}
\usepackage{graphicx}
\usepackage{bm}
\usepackage{color}
\newcommand{\be}{\begin{equation}}
\newcommand{\ee}{\end{equation}}
\newcommand{\bea}{\begin{eqnarray}}
\newcommand{\eea}{\end{eqnarray}}
\newcommand{\bwt}{\begin{widetext}}
\newcommand{\ewt}{\end{widetext}}

\newcommand{\nn}{\nonumber}

\newcommand{\bi}{\begin{itemize}}
\newcommand{\ei}{\end{itemize}}

\newcommand\cC{{\cal C}}
\newcommand\cV{{\cal V}}

\usepackage{setspace}

\begin{document}

\title {Holographic local quench and effective complexity}

\author{Dmitry Ageev, Irina Aref'eva, Andrey Bagrov, Mikhail I. Katsnelson}
\affiliation{Steklov Mathematical Institute, Russian Academy of Sciences, Gubkin str. 8, 119991
Moscow, Russia}
\affiliation{Institute for Molecules and Materials, Radboud University, 6525AJ Nijmegen, \mbox{Netherlands}}
\emailAdd{ageev@mi.ras.ru}
\emailAdd{arefeva@mi.ras.ru}
\emailAdd{a.bagrov@science.ru.nl}
\emailAdd{m.katsnelson@science.ru.nl}

\abstract{
We study the evolution of holographic complexity of pure and mixed states in $1+1$-dimensional conformal field theory following a local quench using both the  ``complexity equals volume'' (CV) and the ``complexity equals action'' (CA) conjectures. We compare the complexity evolution to the evolution of entanglement entropy and entanglement density, discuss the Lloyd computational bound and demonstrate its saturation in certain regimes. We argue that the conjectured holographic complexities exhibit some non-trivial features indicating that they capture important properties of what is expected to be {\it effective} (or physical) complexity.}

\maketitle

\newpage

\section{Introduction}
   
Complexity of a system, a state or a process is one of the most intuitively clear yet
very elusive concepts in our perception of the reality \cite{Jaguar}-\cite{KWK}. It is unlikely that a unique, both universal and practically useful, definition of complexity might exist \cite{Lloyd1}. However sometimes, when dealing with a concrete class of problems (quantum of classical) in a certain area of science, it is of crucial importance to have a particular notion of quantifiable complexity.

Recently, the concept of complexity has attracted a lot of attention in the context of the AdS/CFT correspondence due to its relation to a wide class of the problems concerning holographic entanglement entropy \cite{Ryu:2006bv}-\cite{Swingle:2009bg}, the black hole information paradox, quantum chaos and scrambling \cite{Susskind:2014rva}-\cite{Couch:2016exn}. Concrete ways to calculate complexity of a state in (interacting) quantum field within the holographic framework have been proposed in \cite{Stanford:2014jda}-\cite{Caputa:2017yrh}, and  investigated further in \cite{Carmi:2016wjl}-\cite{Abt:2017pmf}. Two main basic holographic proposals for complexity are the CA (or "complexity=action")\cite{Brown:2015bva}-\cite{Brown:2015lvg},\cite{Lehner:2016vdi} and the CV ("complexity=volume")\cite{Stanford:2014jda},\cite{Alishahiha:2015rta} conjectures. Both have their advantages and arguments in their favor. The CV conjecture relates complexity of a state to volume of a certain codimension-one hypersurface in the bulk, attached to fixed time section of the boundary, while the CA conjecture equates it to the value of gravitational action evaluated in some special region in the bulk. In general, there is no agreement on a preferable covariant formulation of these conjectures for the non-stationary case. Nevertheless, different proposals exist \cite{Susskind:2014rva}-\cite{Abt:2017pmf}. The evolution of volume complexity in the thermofield double state has been studied  in \cite{Brown:2015bva},\cite{Brown:2015lvg},\cite{Couch:2016exn},\cite{Carmi:2017jqz}-\cite{Alishahiha:2018tep}. Dynamics of the CA complexity upon a global quench has also been considered \cite{Alishahiha:2018tep},\cite{Moosa:2017yiz}-\cite{Chen:2018mcc}.

In this paper, we study evolution of both the volume and the action complexity of pure and mixed states (subregions of the total system) in $1+1$-dimensional conformal field theory triggered by a local quench. Roughly speaking, a local quench is a process in which the system gets excited at one point. One example of such a process is local projective measurement resulting in the ``decoherence waves'' \cite{Katsnelson2000}-\cite{Donker2016}. For $1+1$-dimensional CFT, the local quench has been considered in \cite{Calabrese:2007mtj}, and its holographic dual was proposed in \cite{Nozaki:2013wia}. This model has been further generalized in \cite{Caputa:2015waa}-\cite{DeJonckheere:2018pbi}.  Our interest in this setting is two-fold. First, time dependence of quantitative parameters of an out-of-equilibrium system (either quantum or classical) can reveal some hidden physical mechanisms at work, that are not evident from equilibrium considerations. In this regard, our analysis is a direct continuation of the study of a global quench \cite{Alishahiha:2018tep},\cite{Moosa:2017yiz}-\cite{Chen:2018mcc}, and can be interesting on its own. Secondly, and more importantly, a perturbation without translational invariance is very distinct from a global quench as it creates a {\it regular pattern} on top of the static background. When talking about complexity, such a distinction can be decisive.
While the volume and the action conjectures are mainly discussed in the context of (quantum) {\it computational} complexity, and the most random and scrambled states (that is, that cannot be characterized by information contained in the low-order correlation functions) are said to be the most complex, it is tempting to try to employ the holographic intuition to gain a better understanding of complexity of a different type - the {\it effective (physical) complexity} \cite{Jaguar, Lloyd1, Lloyd2, KWK}. Following M. Gell-Mann and S. Lloyd \cite{Lloyd2}, the effective complexity can be defined as an amount of ``non-random'' information that can be non-trivially encoded in {\it regularities} of a system. For example, naturally we do not perceive white noise as a pattern of high informational content, though its complete ``bit-by-bit'' description requires a lot of effort \cite{Jaguar},\cite{Badii_Politi}. This type of complexity that maximizes at the border between order and chaos is the most evading one. And that's what we are talking about when trying to address questions like, - how do biological entities change in the course of evolution \cite{Adami}, in what sense a human being is more complex than a bacteria \cite{Koonin}, or why formally
exponentially hard problems of finding ground-states of many-body quantum systems
can be solved with a relatively simple neural algorithm \cite{Carleo}. Thus, the local quench also serves us as a toy model of creating an ordered structure coexistent with a homogeneous ``randomness''. For simplicity we resort to the most elementary case of a point-like quench model suggested in \cite{Nozaki:2013wia} in pure $AdS_3$ spacetime.

Studying the holographic complexity within this setting, we discover that both the CV and the CA conjectures lead to results that indicate that the holographic complexity have certain features we would expect effective complexity to possess. While the complexity of the systems as a whole steadily increases after the quench, which is along with the original motivation and philosophy of \cite{Brown:2015bva}, time-dependence of the subsystem complexity is much more non-trivial. In particular, we track the evolution of the volume complexity against the entanglement entropy and the integrated absolute entanglement density \cite{Nozaki:2013wia}, that we interpret as a characteristic of inner randomness of the subsystem, and find a regime when complexity rapidly decreases while the system keeps getting more ``random''. On the other hand, we find that saturation of the Lloyd bound for the total system within the CA proposal gets along very well with the intuition behind the theory of self-organized criticalty.

The paper is organized in the following way. In Sec.\ref{sec:setup}, we describe the setup of the holographic local quench and briefly remind the essence of the CV and CA conjectures. Sec.\ref{sec:CV} and Sec.\ref{sec:CA} contain the central technical result of the paper, - the approximate analytical computation of the holographic complexities of the total system and of open subsystems. In these sections we discuss the distinctions in complexity and entanglement dynamics of closed and open systems, and elaborate on the differences between the CV and CA complexities. In Sec.\ref{sec:lloyd} we discuss saturation of the Lloyd bound. We conclude with a discussion, where we try to put the results into a broader context of the problem of physical complexity.

\section{Holographic local quench and complexity in $AdS_3/CFT_2$}\label{sec:setup}

\subsection{Holographic local quench}
A quantum quench of a system is a process triggered by some sudden change of the Hamiltonian. In particular, a local quench can be thought of as a point-like perturbation created at a certain moment of time. In conformal field theory, a possible protocol for making a local quench is to insert a heavy primary operator at certain point in space-time. A simple holographic model of it is naturally given by injection of a massive point-like particle into the bulk of $AdS$ near its boundary \cite{Nozaki:2013wia}. 
Mass $m$ of the injected particle is then related to conformal dimension $h$ of the corresponding heavy operator that triggers the quench
\bea\label{m1}
8  G   m =\frac{24 h}{c},
\eea
where $c$ is the central charge, and $G$ is the gravitational constant. It is handy to introduce notation $M = 8GmL^2    $, where $L$ is the AdS scale defined below.

Consider the Poincare patch of $AdS_3$ space-time:
  \bea\label{pads}
ds^2=\frac{L^2}{z^2}(-dt^2+dx^2+dz^2), \,\, z>0.
 \eea
Action of a massive particle moving along $x=0$ line is given by
\bea\label{part}
S=-m L\int  \frac{\sqrt{1-\dot z(t)^2}}{z(t)}dt
\eea
Its trajectory in the bulk is simply
\bea\label{eq:traj}
&&z(t)=\sqrt{t^2+\alpha^2},\\
&&x(t)=0, \nonumber
\eea
where $\alpha$ is the initial distance from the particle to the boundary. It can be shown that \eqref{eq:traj} remains valid not only in the probe limit, but also when the particle deforms the metric of $AdS_3$. To compute backreaction of the particle following worldline \eqref{eq:traj} on metric \eqref{pads}, we use the following trick. First, we consider a static point-like particle of the same mass in global $AdS_3$, with the resulting metric in global coordinates
\be \label{metr1}
ds^2=-d\tau^2
   \left(L^2-M+R^2\right)+R^2 d\phi^2+\frac{L^2 dR^2 }{L^2-M+R^2},
\ee
where the particle sits at $R=0$.  This metric describes a conical defect for $M<L^2$, and the BTZ black hole for $M>L^2$. 

Then we apply a coordinate transformation that relates (half of) the global $AdS$ and the Poincare patch
\bea\label{map}
&&\phi =\arctan\left(\frac{2 \alpha  x}{\alpha ^2+t^2-x^2-z^2}\right)\\ \nonumber
&&\tau =\arctan\left(\frac{2 \alpha  t}{\alpha ^2-t^2+x^2+z^2}\right)\\ \nonumber
&&R=\frac{\sqrt{\alpha ^4+2 \alpha ^2 \left(t^2+x^2-z^2\right)+\left(-t^2+x^2+z^2\right)^2}}{2 \alpha  z}.
\eea
We find that \eqref{map} maps the wordline of a static particle ($R=0$) onto trajectory of an infalling particle, $z=\sqrt{\alpha^2+t^2}$, in the Poincare patch, and obtain a complicated time-dependent metric $g$ (see App.\ref{appA} for its explicit form) which is the holographic dual of the local quench. When $M=0$, transformation \eqref{map} maps \eqref{metr1} to \eqref{pads} for arbitrary $\alpha$. The boundary stress-energy tensor dual to metric \eqref{eq:long_metric} matches the result of the corresponding field theory calculation \cite{Calabrese:2007mtj}-\cite{Nozaki:2013wia}, providing an additional evidence in favor of reliability of the model. As was shown in \cite{Nozaki:2013wia}, the quench leads to formation of a pair of entangled ``solitons'' on the boundary, propagating away from the point of perturbation. On the boundary, $\alpha$ defines the size of the perturbation and the width of the solitons. The case of $\alpha \rightarrow 0$ corresponds to the vanishing solitons size and infinite energy density injection, or, in other words, to the quench exciting arbitrarily high frequency modes.

The relation between the bulk particle energy and the conformal dimension of perturbing operator in this model is
\bea\label{m2}
E=\frac{m}{\alpha}L=3\frac{h}{\alpha c G}=2\frac{h}{\alpha},
\eea
where we used the relation between the gravitational constant and the central charge 
\bea\label{m3}
G=\frac{3L}{2c}.
\eea
For simplicity, hereinafter we take $L=1$.

\subsection{CV and CA conjectures}

There are two main conjectures relating complexity of a state in conformal field theory to the dual bulk quantities, -- the ``complexity equals volume'' (CV) and the ``complexity equals action'' (CA) dualities.  
 
The CV duality defines complexity $\cC_{\cV}(t)$ of the boundary state at time $t$ in terms of a codimension-one bulk hypersurface $B$ attached to the fixed time slice of the boundary
 \bea
\cC_{\cV}(\Sigma)=\frac{\cV(B)}{G L},
 \eea
where $\cV(B)$ is volume of this hypersurface, $G$ is the gravitational constant, and $L$ is the characteristic scale of the bulk geometry (for example the $AdS$ radius). There are different ways to motivate this definition. Originally it comes from the observation that in an out-of-equilibrium holographic setting volume of the Einstein-Rosen bridge behind the black hole horizon keeps growing long after the system reached local thermal equilibrium. Another motivation emerges from the tensor network interpretation of the AdS/CFT correspondence (see numerical experiments in \cite{Abt:2017pmf} supporting this relation). Another interesting aspect of this is the relation between the bulk volume and the Fisher information metric and bulk entanglement \cite{Banerjee:2017qti}.

The CA duality equates complexity to the value of certain gravitational action (\cite{Parattu:2015gga}-\cite{Lehner:2016vdi}) evaluated on the Wheeler-DeWitt (WDW) patch
 \bea
\cC_{A}(t)=\frac{S({\cal W})}{\pi },
 \eea
where WDW patch $\cal W$ is the bulk domain of dependence of any Cauchy surface
 which asymptotically approaches the fixed time slice of the boundary. This conjecture was proposed on the basis of analysis of the worldvolume behind the horizon of an eternal black hole corresponding to the thermofield double state in the boundary field theory. The constructive way to define the WDW patch is to select bulk region bounded by null rays sent from the boundary at time moment $t$ in both directions. 

The generalization of these conjectures onto the open subsystem case is not straightforward. Since the
general idea of the ``complexity equals volume'' conjecture is to compute the volume of some space-time subregion which asymptotically approaches given fixed time boundary slice, the most natural way to define the subsystem complexity is via the bulk volume of the region bounded by the corresponding Ryu-Takayanagi surface in the static case, and by the Hubeny-Rangamani-Takayanagi (HRT) in the dynamical case \cite{Alishahiha:2015rta}, \cite{Carmi:2016wjl}.
A more sophisticated approach to this has been suggested in \cite{Carmi:2016wjl}, but in this paper we stick to the simpler version.
In the same paper \cite{Carmi:2016wjl}, a generalization of the CA duality for subregions has been proposed. The idea is to evaluate the gravitational action not on the whole WDW patch, but on its intersection with the entanglement wedge corresponding to the subregion of interest. That is the strategy we follow here when discuss the action complexity.

The two conjectures have some seeming advantages and disadvantages. In particular, formulation of the CV conjecture requires existence of some scale $L$, and involving such an arbitrary (from the boundary point of view) parameter is undesirable \cite{Brown:2015lvg}. The CA conjecture does not require such a parameter, but it arises from some boundary counter terms (for example, see \cite{Carmi:2017jqz}). On the other hand, the volume complexity can be naturally related to the tensor network representation of the AdS/CFT duality \cite{Abt:2017pmf}.

Before we turn to the non-equilibrium setting, it is instructive to consider complexity of static metric \eqref{metr1}. The action complexity can be expressed as (see App. \ref{sec:WdW} for the detailed derivation):
\be 
\cC_{I}=\frac{c}{6} \sqrt{1-24h/c}-\frac{c}{3\pi}\frac{r_m}{L}+\frac{S_p}{\pi},
\ee
where $r_m$ is the boundary cutoff ($r_m \rightarrow \infty$), and $S_p$ is the static particle action.

The volume complexity (volume of constant time slice in \eqref{metr1}) is
\be 
\frac{\cC_{\cV}}{2\pi}=\frac{2c}{3}\left(\frac{r_m}{L}-\sqrt{1-\frac{24h}{c}}\right)
\ee

\section{Volume complexity}\label{sec:CV}
To compute the holographic volume complexity of a boundary subregion one has to construct the HRT surface homologous to the subregion \cite{Alishahiha:2015rta}-\cite{Carmi:2016wjl}, which is a covariant generalization of the static Ryu-Takayanagi minimal surface. In a generic case of time-dependent background, this might be a non-trivial task. However, in the case of point-like perturbation of $AdS_3$ we can readily proceed. In principle, the HRT surface then can be even calculated exactly, but for simplicity we will use the perturbative method \cite{Nozaki:2013wia}, originally proposed for computing the holographic entanglement entropy. The main assumption of the method is that for small mass of the perturbation $M$ the HRT surface can be approximated by that of the unperturbed spacetime, but the induced metric on it should inherit from the metric of the deformed spacetime. Calculation of the entanglement entropy and the volume complexity can then be carried out within this approximation.

It was shown both for $AdS_3$ \cite{Nozaki:2013wia} and for higher bulk dimensions \cite{Jahn:2017xsg} that this method in fact gives accurate results, very close to the exact answer in a wide range of parameters.

\subsection{Total system complexity \label{Sec:FullCV}}

Before computing subregion complexity, we will analyze dynamics of the total system complexity after the local quench. The volume then is the (renormalized) volume of the constant time slice of the whole spacetime. We denote the metric on the slice $\Sigma_t$. Still, it is convenient to represent such a slice as a domain under a HRT surface stretched between infinitely distant boundary points.

\begin{figure}[h!]
\centering
\includegraphics[width=7.5cm]{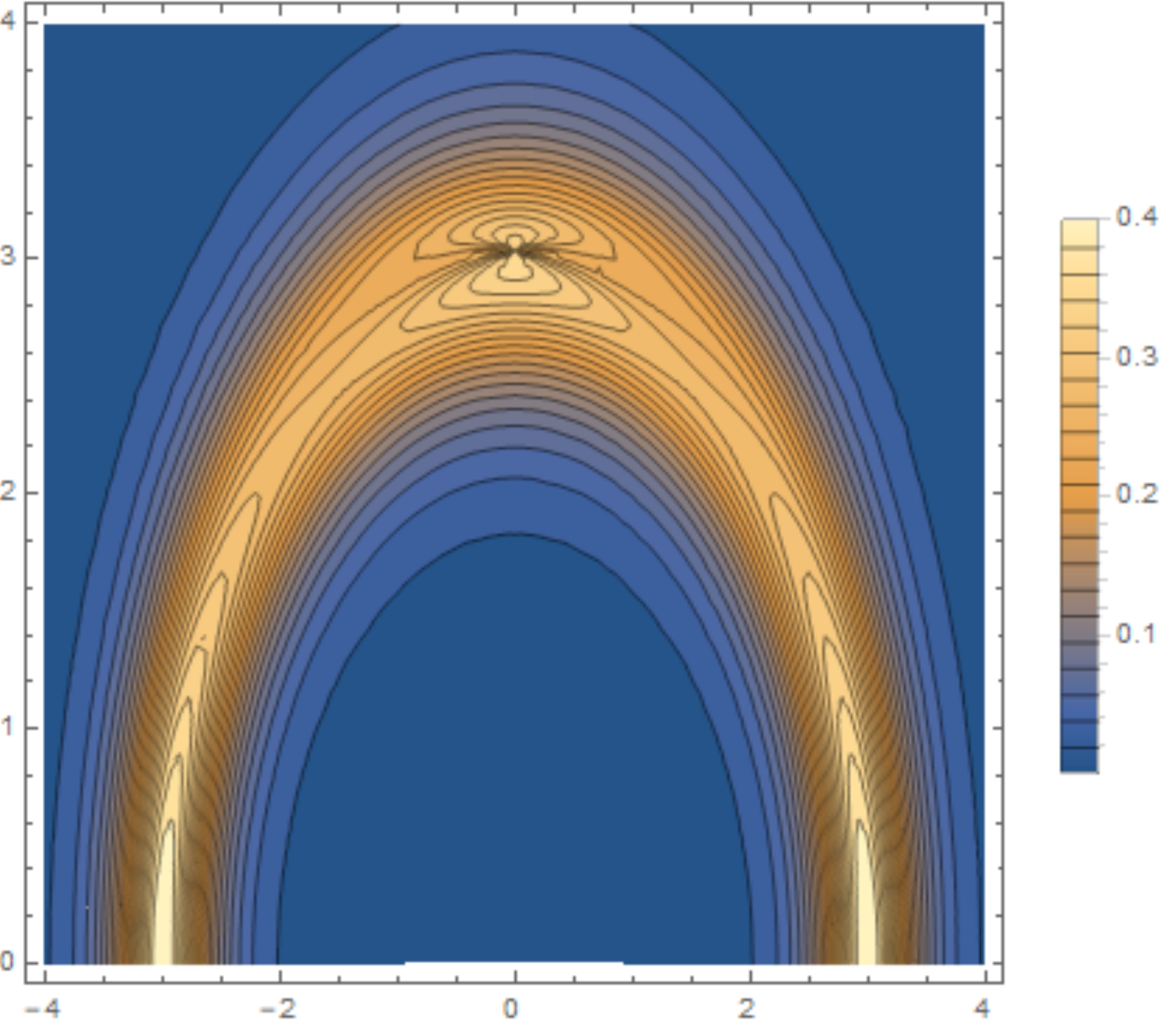}
\includegraphics[width=7.5cm]{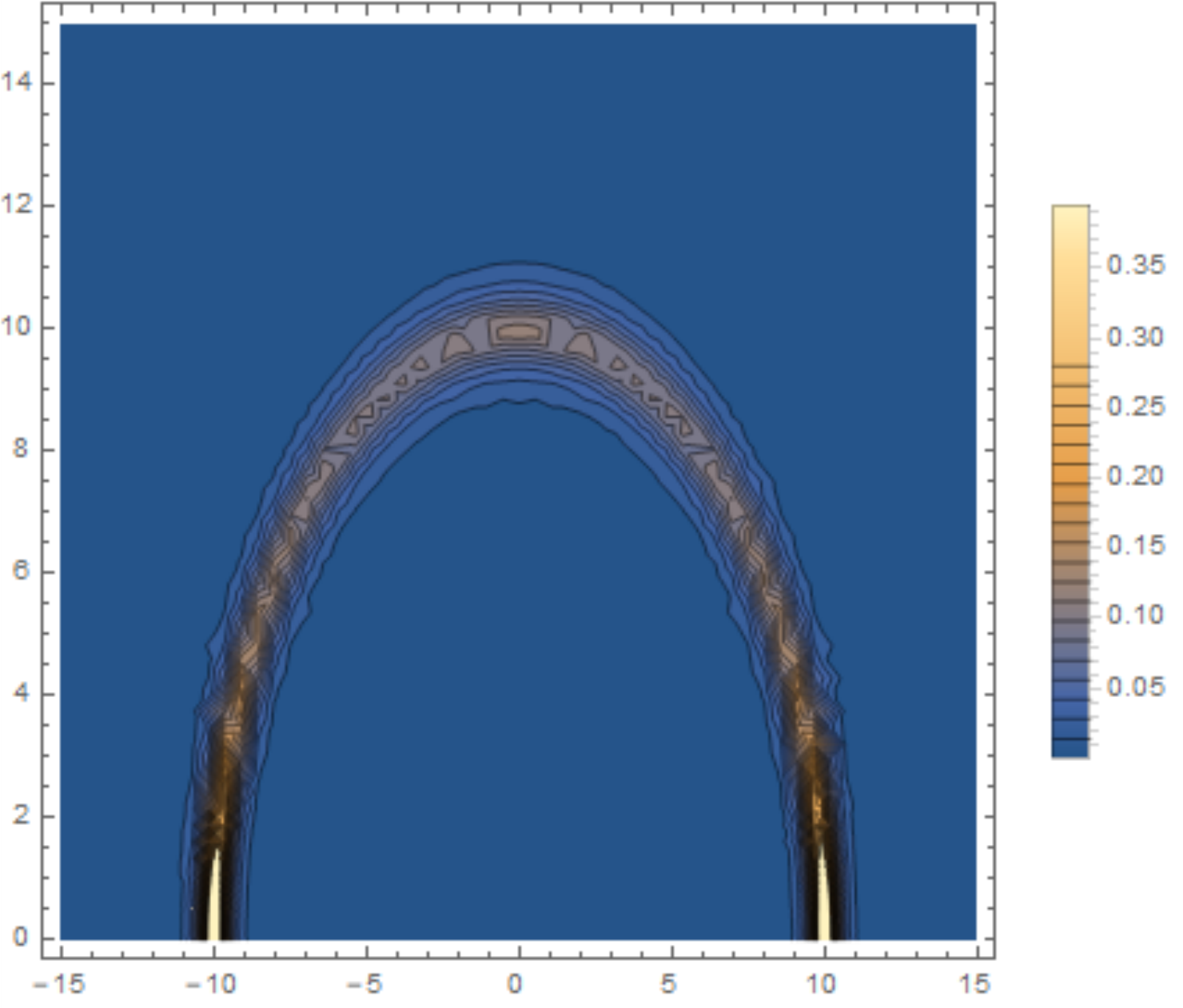}
 \caption{The density of the renormalized volume form $\Sigma$ for $M=0.2$ and $\alpha=0.5$. The left plot corresponds to the time moment $t=3$, and the right one to $t=10$.}
 \label{fig:sigmat}
\end{figure}

To get some intuition on how the complexity and the entanglement entropy respond to the metric perturbation caused by a massive particle, we shall study how the constant time slice volume form $\sqrt{\det \Sigma_t}$ (explicit form of the metric is \eqref{eq:long_metric}) evolves in time. Let us define renormalized quantity $\Sigma=\Sigma(t,x,z,\alpha,M)$ as
\bea\label{DV}
\Sigma=\sqrt{\det \Sigma_t}-\frac{1}{z^2}.
\eea
We plot this quantity in Fig.\ref{fig:sigmat} for different moments of time. The massive source is moving away from the boundary into the bulk along $x=0$ line, and induces an annulus-shaped trace of metric perturbation behind it which is expanding in time. After the system has evolved enough, we can see that near the boundary the density perturbation of $\Sigma$ is much sharper than in the bulk (right part of the plot). 
Deviation of the volume complexity from its equilibrium value is defined as $\Sigma$ integrated over the whole time slice
\bea\label{CS}
\Delta{\cal C}(t)=\frac{2c}{3}\int\limits_{z>0}\int\limits_{x\in {R}} \Sigma dx dz,
\eea
where $c$ is the central charge of the boundary CFT.
For small M ($M\ll L^2$) nd early times ($t\rightarrow 0$)  one obtains quadratic growth of the system complexity: 
\bea
\Delta {\cal C}(t)\approx 16 \pi h (1 +\frac{t^2}{2\alpha^2}),
\eea
where we used the expansion from App.\ref{pert} and formulas \eqref{m1}-\eqref{m3}. 
Note that the leading term in this expansion is of zero's order in time, and depends only on conformal dimension $h$ (equivalently, mass $M$). At later times $\Delta {\cal C}$ must be computed numerically. In this section and in the next one, we use the entanglement entropy $\Delta S$ rescaled by the factor $1/4G$ and the complexity (calculated by formula \eqref{CS}) rescaled by $3/2c$.

\begin{figure}[t!]
\centering
\includegraphics[width=7.5cm]{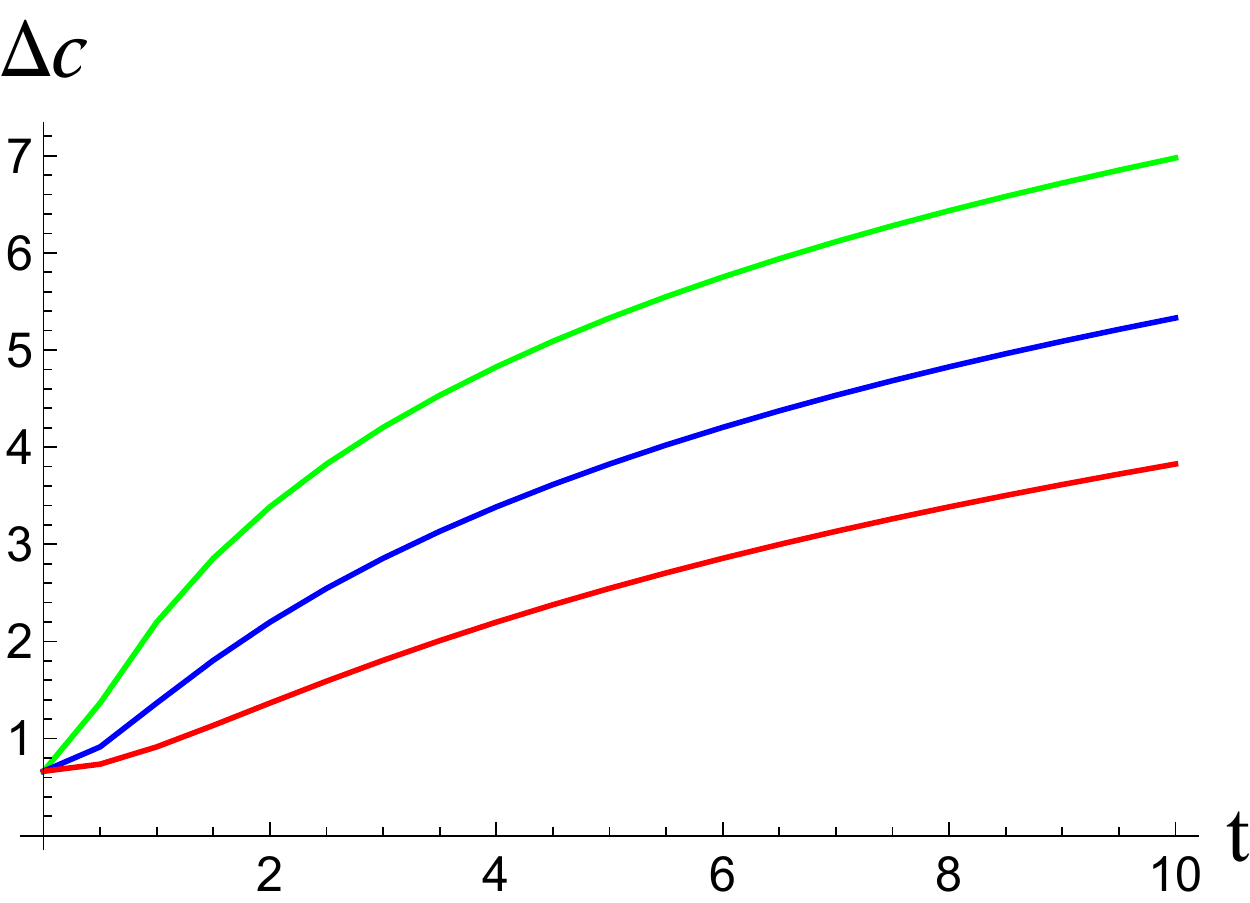}
\includegraphics[width=7.5cm]{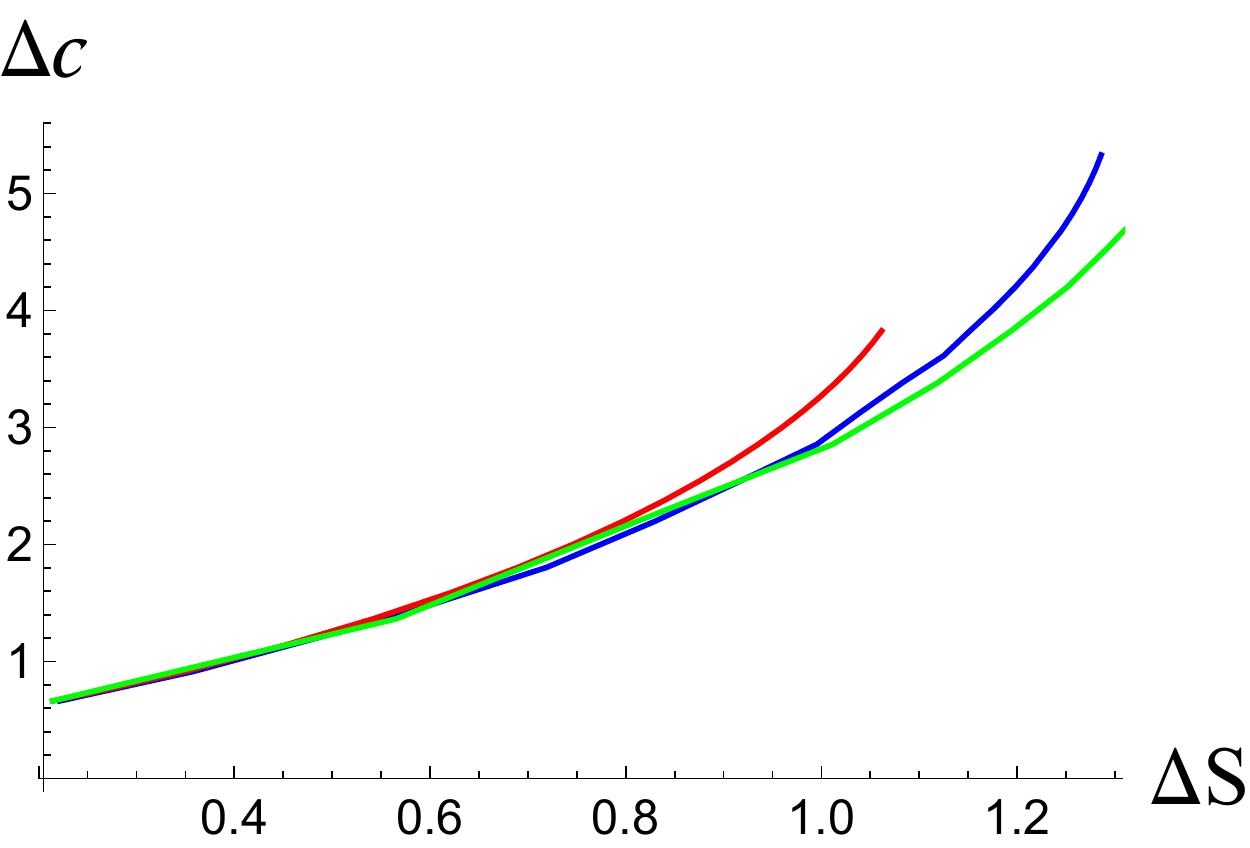}
 \caption{On the left plot, we show the time dependence of the (rescaled) complexity $\Delta \cC$ of the excitation. On the right plot, we present the dependence $\Delta \cC(t)$ on (rescaled) $\Delta S(t)$ as time changes from $t=0$ to $t=10$. The green curves correspond to $\alpha=0.25$, the blue ones -- to $\alpha=0.5$, and the red ones -- to $\alpha=1$.}
 \label{fig:cwhole}
\end{figure}

Since we deal with the total system in a pure state, to define the entanglement entropy we consider its half-space bipartition into two semi-infinite parts along $x=0$ line. It is known that for a local quench the entanglement entropy of such an infinite subsystem demonstrates logarithmic growth as $t \rightarrow \infty$ \cite{Calabrese:2007mtj}-\cite{Nozaki:2013wia}. In Fig.\ref{fig:cwhole} we plot the evolution of $\Delta \cC $ for the total system, and compare it to the evolution of entanglement entropy of this bipartition.

We see that for some time right after the quench the complexity of the excited state is proportional to the entanglement entropy, and is independent of the quench strength $\alpha$. In \cite{Nozaki:2013wia}, the amount of information carried by a local excitation was estimated to be $E\times\alpha \approx 2h$, which means that during the early time evolution the volume complexity is proportional to the information of the excitation. This dependence is followed by a power law at later times, which is non-universal in $\alpha$. 

\subsection{Subsystem complexity}
Our next goal is to study the evolution of complexity for a system that was initially in a mixed state, and we will see that it exhibits some distinctive features as compared to the system in a pure state. To achieve that, we consider a subregion of the boundary, which is taken to be a single interval $x \in (-\ell,\ell)$.

\begin{figure}[t!]
\centering
\includegraphics[width=7.5cm]{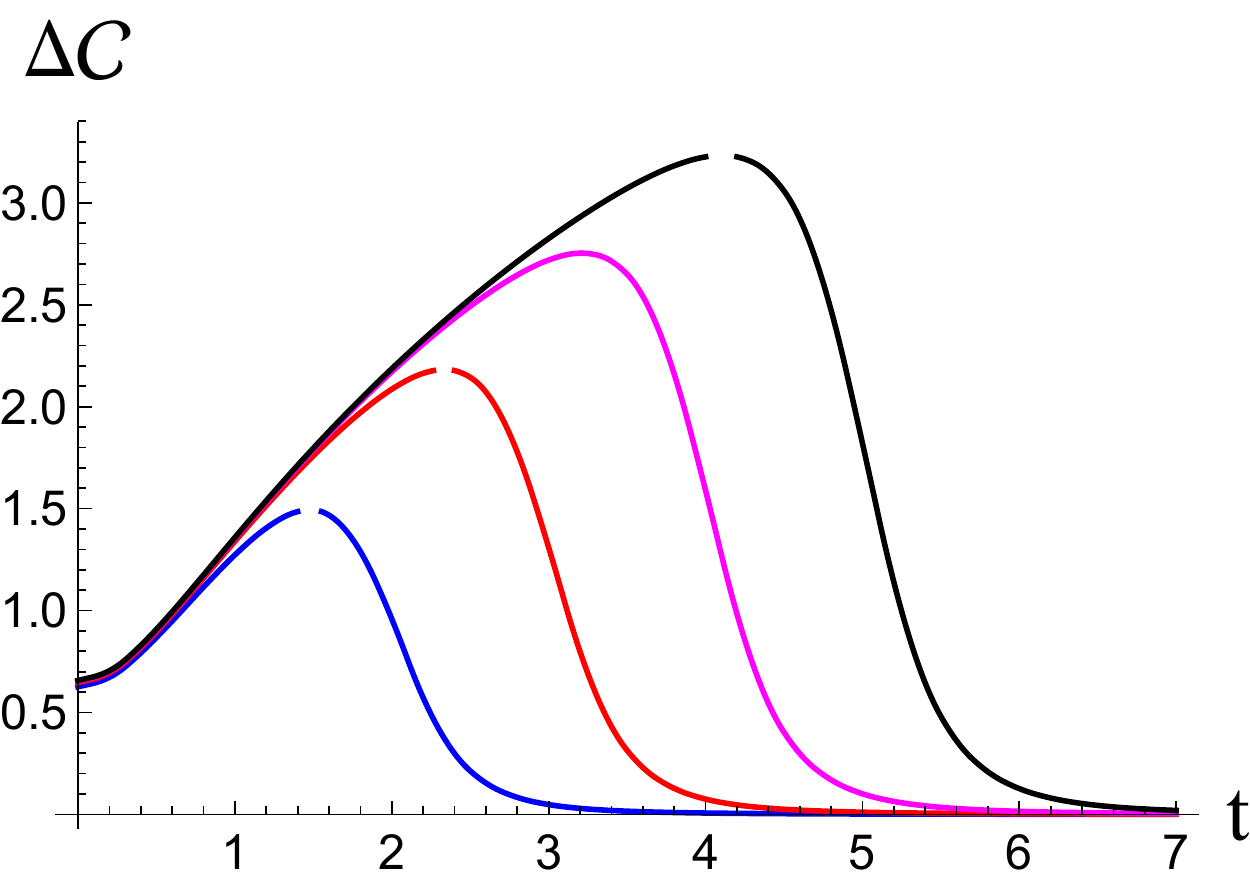}
\includegraphics[width=7.5cm]{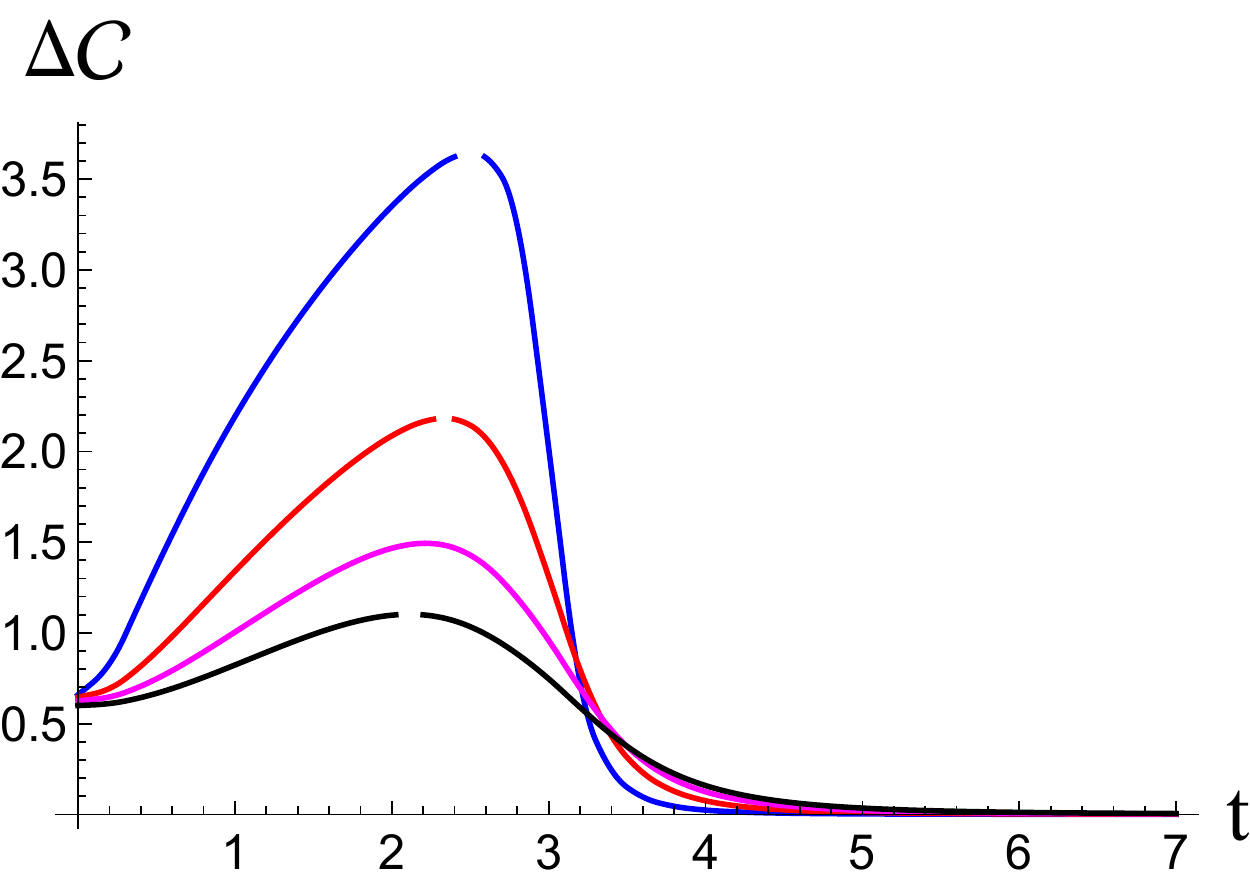}
 \caption{On the left plot we show the time dependence of the (rescaled) complexity of the excitation $\Delta \cC$ for interval $x\in(-\ell,\ell)$ and for $\alpha=0.5$. Different curves correspond to different $\ell=2,3,4,5$ for down to top.  On the right plot the same for fixed $\ell=2$ and $\alpha=0.25,0.5,0.75,1$ from down to top.}
 \label{fig:csell}
\end{figure}

First, assume the quench is sharp, $\alpha<\ell$. Right after the quench, for a short time the complexity demonstrates quadratic growth (as in the case of total system, \ref{Sec:FullCV}), which is followed at larger times by the regime of linear growth that continues almost all the way up to the point where the complexity becomes maximal, see Fig.\ref{fig:csell}. After passing the maximum, the complexity smoothly but quickly returns to its equilibrium value. 

As we expressed in the introduction, our primary interest is to see whether the proposed conjectures of holographic complexity allow to gain some understanding of the concept of effective complexity. In order to attain this, we need to contrast the evolution of complexity evolution against some measure of ``randomness'' (or typicality) of a state. In the classical case the thermodynamical entropy would be an obvious choice of such a measure. However, the quantum case is more peculiar. The entanglement entropy is not a good candidate for this role, as it is rather a characteristic of the interface between the (sub)system and the environment, and does not tell much about the inner structure of the state. Still, it is a natural departing point for our analysis, and for a moment we will focus on it.

\begin{figure}[t!]
\centering
\includegraphics[width=7.5cm]{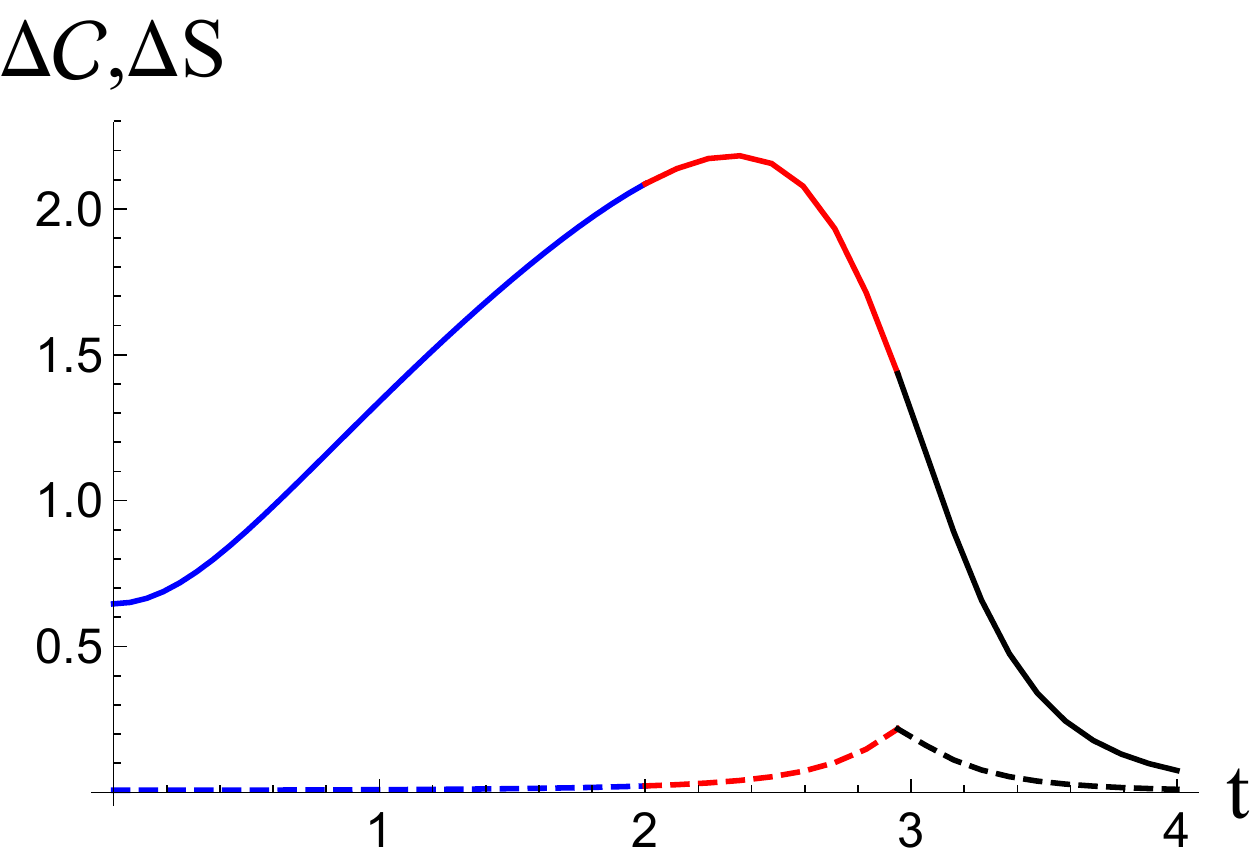}
\includegraphics[width=7.5cm]{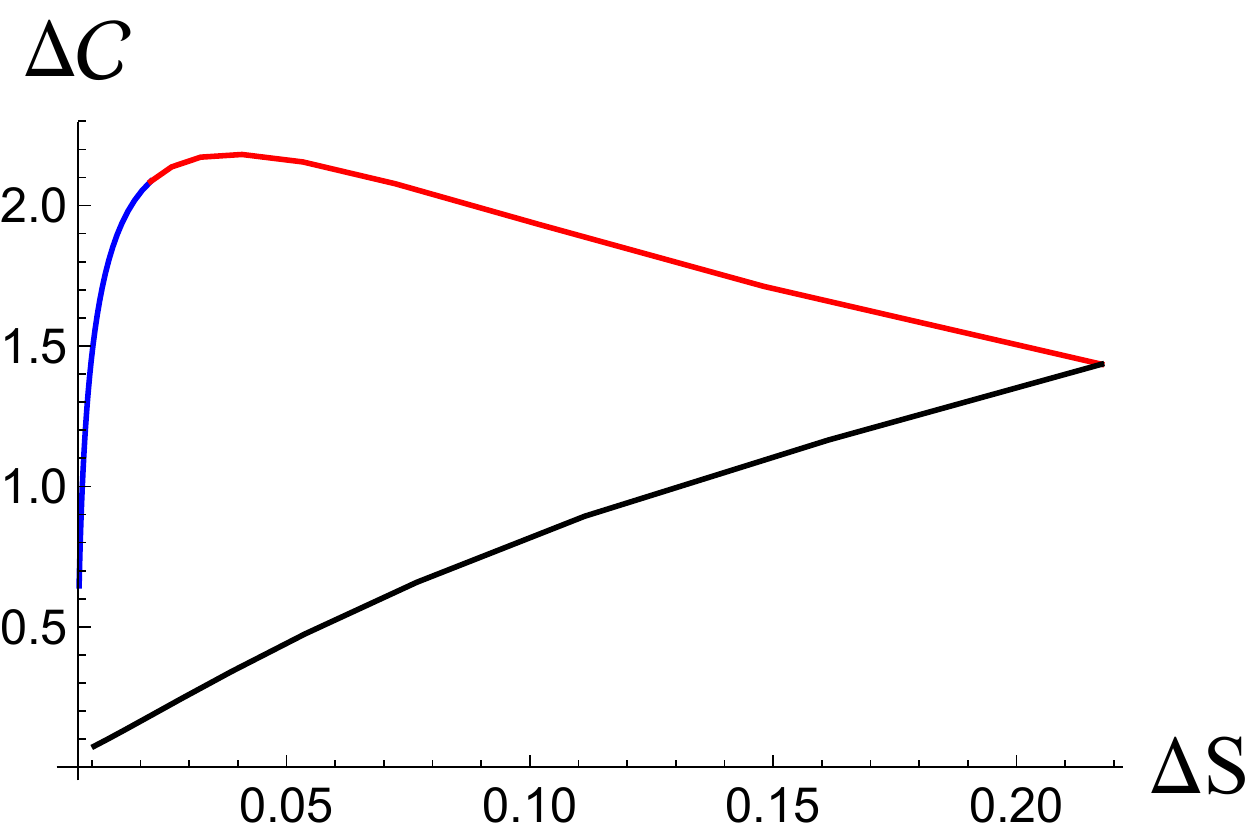}
 \caption{On the left plot we present the time dependence of the (rescaled) complexity (solid curve) and the (rescaled) entanglement entropy (dashed curve) of $x\in(-3,3)$ interval for the same parameters. The right one is a parametric plot where we show the complexity $\Delta \cC$ of the quenched subsystem as a function of (evolving in time) entropy. The red, the blue and the black curves here correspond to the same time intervals as on the left plot.}
 \label{fig:CVell}
\end{figure}

Evolution of the entanglement entropy can be computed in a straightforward way (\cite{Nozaki:2013wia}, refer to App.\ref{pert} for the details), and one can see that it firstly increases demonstrating a very sharp peak around $t \sim \sqrt{\ell^2+\alpha^2}$, and then rapidly approaches the original equilibrium value, and, what is more, its dependence on time is almost symmetric around the maximization point.
To compare the patterns of $\Delta S$ and $\Delta C$ thermalization, we put their time dependences on the same plot, Fig.\ref{fig:CVell}. First of all, we have to note that the two quantities reach their maxima at different moments of time. Between the two periods when both quantities grow or both decrease, there is a period when complexity has already reached its maximal value and decreases, while entropy keeps growing. Though, as we pointed out, the entanglement entropy is not a perfect characteristic of randomness of the internal structure of a state, this type of behavior can be regarded as a sign of what one expects to see for the effective complexity rather than for complexity of the exact description of a state.

Let us take a closer look at the dependence of the complexity $\Delta C$ on the entropy $\Delta S$. The three regimes of evolution are then mapped onto three parts of parametric plot Fig. \ref{fig:CVell} (right plot):
\begin{itemize}  
\item The initial stage of evolution (the blue curve) is characterized by a relatively weak growth of the entanglement entropy, while the complexity is growing pretty fast. This is in accordance with the intuitive physical interpretation that the quenched state is very different from the ground state, and as a result the complexity of its preparation is very high. On the other hand, the sharp local quench creates a pair of particles that, being highly entangled with each other, do not contribute to the entanglement between the subsystem and the environment. It is worthy to note that the growth of complexity correlates with the increase of distance between the particles. A possible interpretation of this relation is that the ``Bell pair'' moving from the center to the boundary of the interval leaves an {\it entanglement trace} behind it and causes a state restructuring, making it more complex.
\item  The second stage of the evolution (the red curve) is the most interesting one. The entanglement entropy keeps growing, while the complexity is decreasing, and the dependence of $\Delta C$ on $\Delta S$ is nearly linear. At this stage the pair approaches the boundary of the interval, and thus the entropy rapidly maximizes - when the solitons are right on the boundary, their mutual entanglement contributes to the entanglement between the subsystem and its complement. The reason for complexity to decrease is less evident. Also it is interesting to mention that, while the entanglement entropy varies quite strongly, the complexity mildly changes with respect to entanglement. This implies that, when certain level of entanglement is achieved, the subsequent restructuring of the state is easier to perform than when depart from the ground state.
\item At the third stage both the complexity and the entanglement entropy decay, and are almost linearly proportional to each other. The subsystem relaxes to its equilibrium state, and all the non-trivial perturbations escape to the environment.
\end{itemize}

This kind of behavior is already interesting on its own, but to reinforce the claim that the volume complexity of an open system is somewhat similar to the intuitively understood effective complexity, we need to suggest a more reliable measure of the ``inner randomness'' of a state. We shall build this notion on the basis of entanglement density $n(t,\xi,\ell)$ introduced in \cite{Nozaki:2013wia} as a density of ``Bell pairs'' of given radius $\ell$ centered around $\xi$:
\begin{equation}
n(t,\xi,\ell)=\frac14 \frac{\partial^2 S(t,\xi,\ell)}{\partial \xi^2}-\frac{\partial^2 S(t,\xi,\ell)}{\partial \ell^2}
\end{equation}
The higher the deviation of inner entanglement rate from the vacuum one is, the more ``random'' state of the subsystem we claim to be. Thus we define the integrated entropy density
\begin{equation} \label{eq:integrated_density}
\mathcal{N}(t) = \int\limits_{-L}^{L}d\xi \int\limits_{0}^{L}d\ell |n(t,\xi,\ell)|,
\end{equation}
where absolute value of $n(t,\xi,\ell)$ is taken since we are interested in the total deviation from the reference vacuum state. This characteristic is inspired both by the concept of thermodynamical Gibbs entropy in classical statistical physics, and by the theory of complex networks. 
While these analogies should not be taken too literally, it is instructive to elaborate on them.
The Gibbs entropy of a classical system defines how many different microscopic states can be drawn from an ensemble of systems in the same thermodynamical state. In quantum theory, a wavefunction defines the statistics of outcomes of measurements performed on an ensemble of identical states. For the sake of lucidity, imagine a quantum spin chain, and a classical detector that measures projection of each spin on the given direction (which may be different for different spins). If the chain is in a product state, we can always tune the local (on-site) basis of the detector in such a way that the measurement will give exactly the same outcome when repeated. On the other hand, if there are Bell pairs within the many-body state, the number of possible outcomes will grow exponentially with the number of Bell pairs, pretty much as the number of microscopic states in a classical ensemble grows exponentially with entropy. \footnote{We would like to stress out one more time that this analogy should be considered as an ideological rather than a formal one. If the inner entanglement is multipartite, its structure can be represented by Bell pairs only approximately (at best). Another issue is related to the fact that here we talk about possible deviations not from a products state, as in the spin chain example, but from the entangled CFT vacuum. Still, we think $\mathcal{N}$ is a legitimate measure of randomness of a state, though not a unique one.} 

\begin{figure}[t!]
\centering
\includegraphics[width=7.5cm]{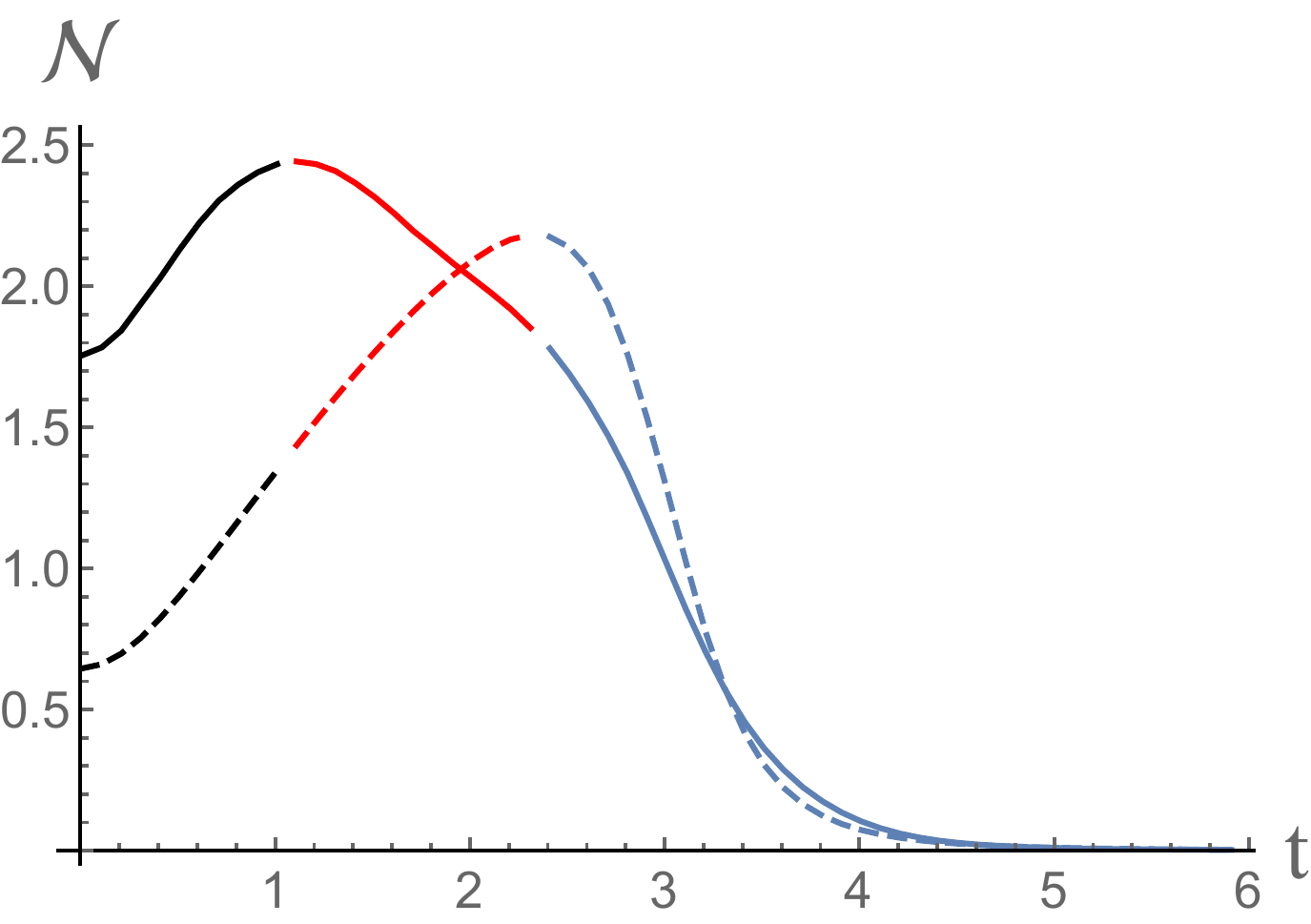}
\includegraphics[width=7.5cm]{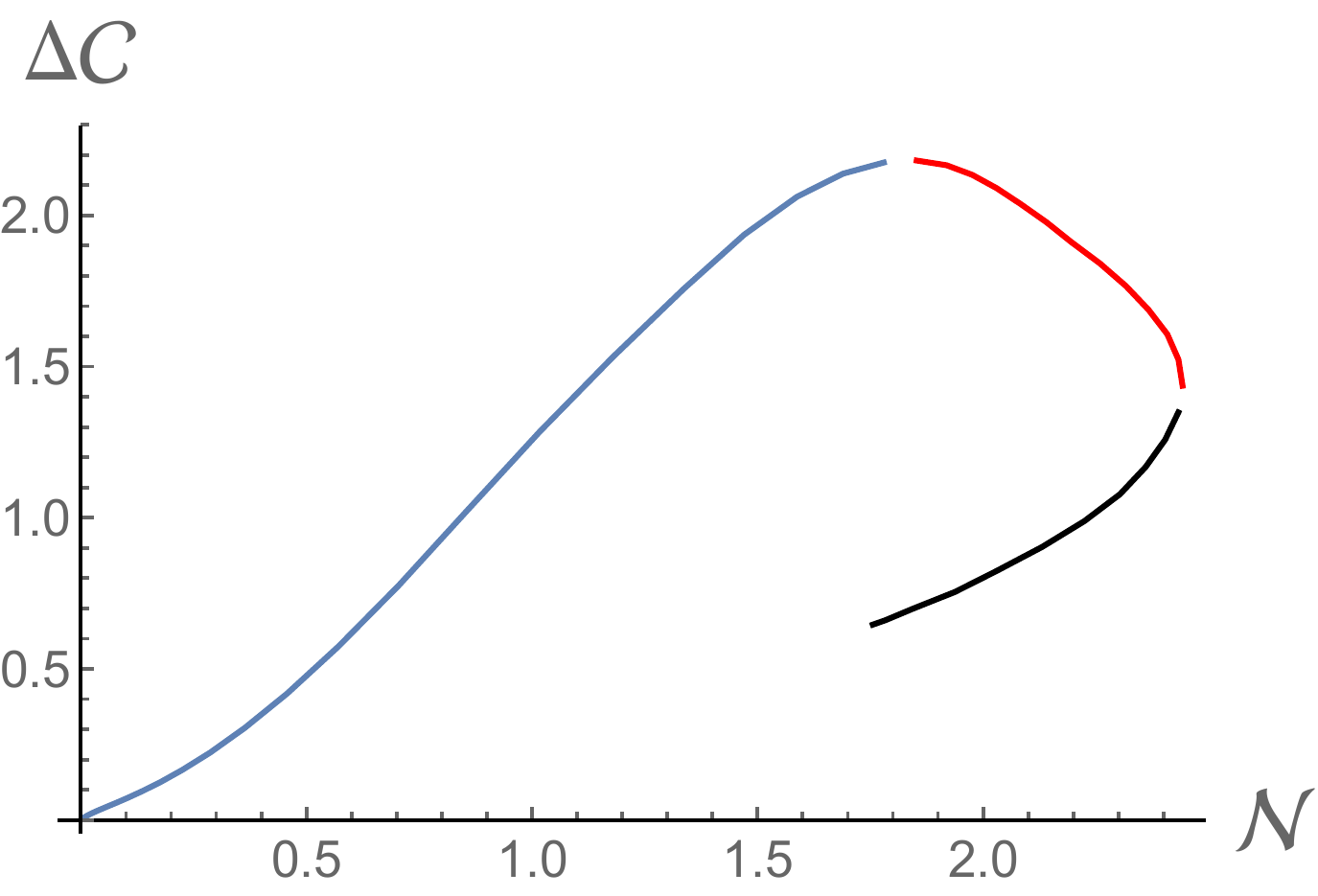}
 \caption{Left plot: time dependence of the integrated entanglement density $\mathcal{N}$ (solid) and the volume complexity (dashed) of the interval $x\in (-3,3)$ at $M=0.2,\,\alpha=0.5$. Right plot: the relation between the (rescaled) subsystem volume complexity $\Delta C$ and $\mathcal{N}$.}
 \label{fig:csell-dens}
\end{figure}
Another way to look at \eqref{eq:integrated_density} is to draw an analogy with the concept of complex networks. It has been proposed \cite{CarrNetwork} that a state of a lattice many-body quantum system can be represented as a network of nodes (lattice sites) connected by links weighted with the corresponding intersite mutual information. The degree of randomness is then associated with the clustering of the network, which is minimal near the quantum critical point \cite{CarrNetwork},\cite{CarrNetwork2}. $\mathcal N$ is a direct analogy of that, with the reserve that here we deal with a continuum limit, and the links are weighted with deviations of entanglement entropy from the vacuum state instead of mutual information.
Again, as for the entanglement entropy, we find a regime where complexity depends on $\mathcal N$ non-monotonously, Fig.\ref{fig:csell-dens}, giving yet another argument in favor of the statement that the holographic volume complexity can be regarded as a certain version of effective complexity.

\section{Action complexity}\label{sec:CA}
\subsection{Total system complexity}
Now we turn to the description of the complexity evolution following the local quench assuming the CA conjecture. The latter relates the complexity of boundary system at time $t$ to the bulk action evaluated within a certain space-time region called Wheeler-DeWitt patch (WDW), that is bounded by the set of null rays emanating from the constant time slice of the boundary. The gravitational action to be evaluated consists of three types of terms: the volume term, which in our case is simply proportional to the volume of the WDW patch (because even in the presence of a massive particle the spacetime scalar curvature is constant), the terms coming from the joints of the WDW patch and the entanglement wedge (where the null rays intersect), and terms that have to be introduced in order to restore the diffeomorphism invariance \cite{Lehner:2016vdi}. The contributions of different terms cause some additional divergences, and in the case of finite interval complexity it is not clear yet how to get a handle on them \cite{Carmi:2016wjl}. Therefore we limit ourselves by the analysis of possible effects related to the matter (massive particle) term. In another words, we work in the probe approximation(note, that probe approximation for the string has been considered
in the context of complexity in \cite{Nagasaki:2017kqe}). Focusing solely on this contribution, we obtain an estimate on $\Delta \cC (t)$ (the volume contribution in our approximation is nearly constant for all $t$, and can be ignored) and elaborate on the main differences with the volume complexity behavior. Full detailed analysis concerning the action complexity evolution upon the local quench will be presented elsewhere. Note that the $AdS_{3}$ case is quite special, and the point-like massive particle has a stronger effect on the background than in higher dimensions. However, the local quench protocol based on the falling particle model is pretty universal, - and the qualitative picture remains similar \cite{Nozaki:2013wia}, allowing to generalize the discussion below onto higher dimensional systems.

\begin{figure}[h!]
\centering
a)\,\,\includegraphics[width=5.5cm]{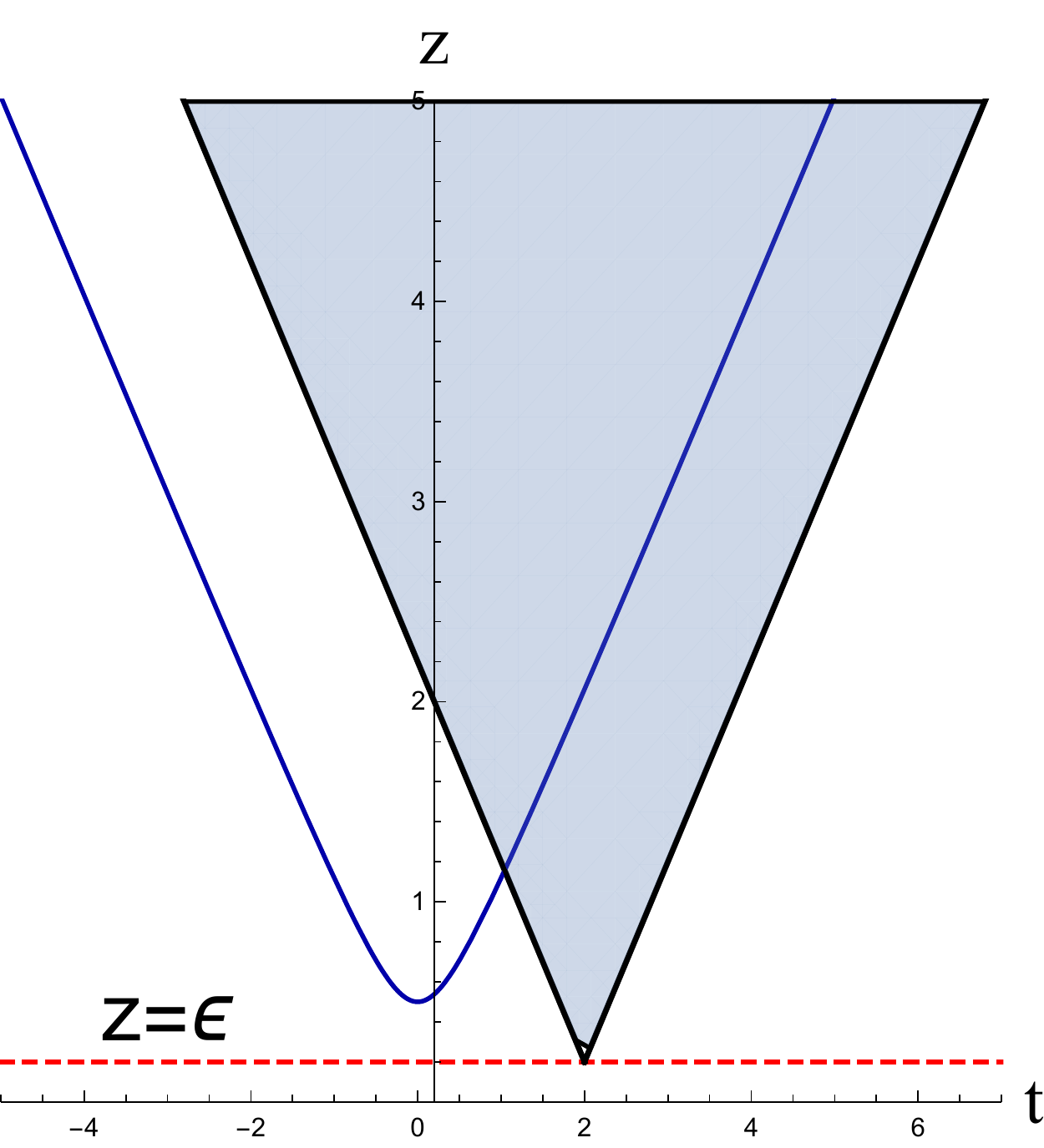}
b)\,\,\includegraphics[width=5.5cm]{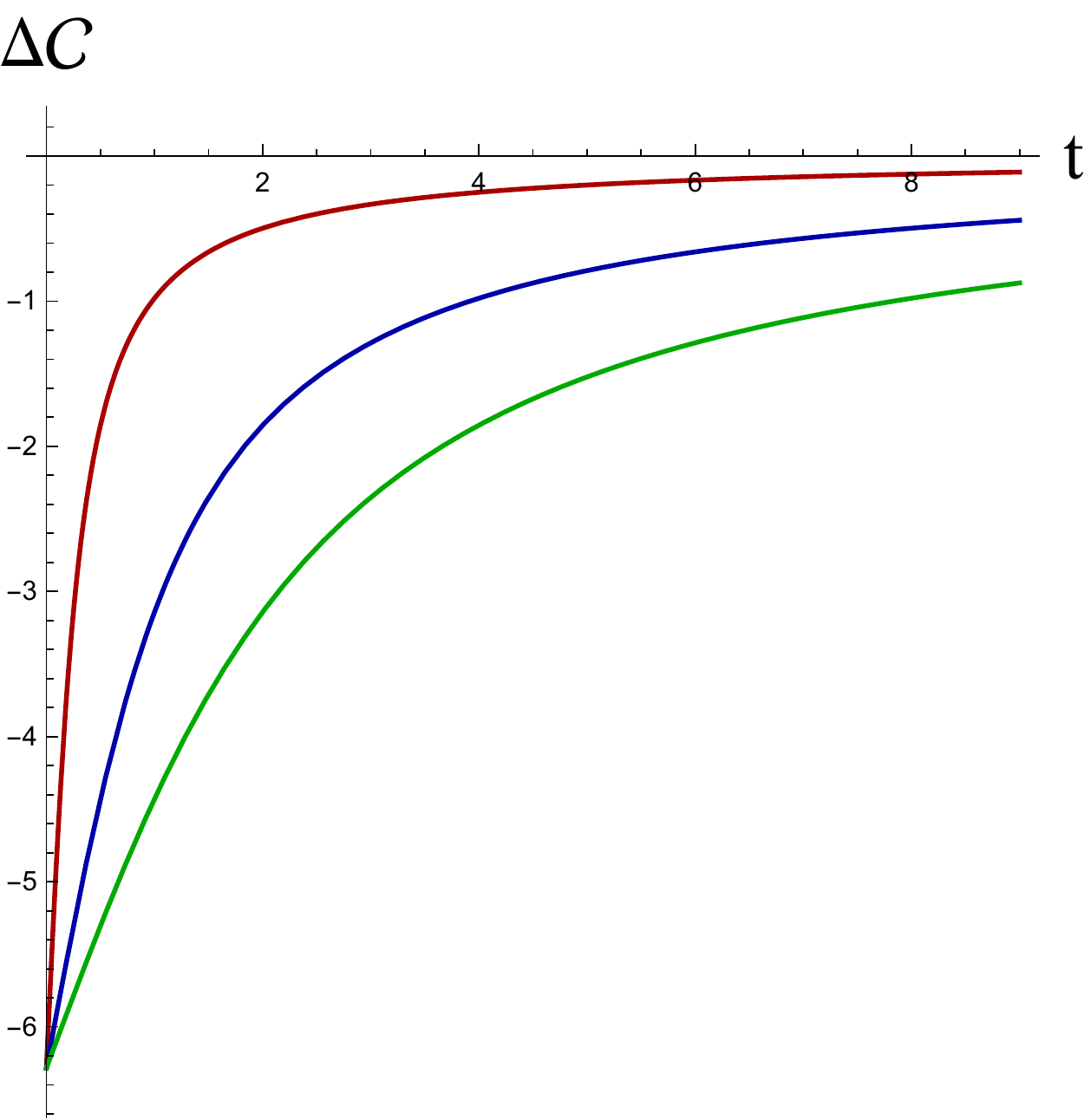}
 \caption{In the left plot, we present the Wheeler-DeWitt patch corresponding to the holographic local quench setup. The blue curve is the massive particle trajectory, the red dashed line is the UV regulated boundary, and the black solid lines are the boundaries of the WDW patch. In the right plot, the time evolution of the action complexity is shown, $\alpha=0.5,1,2$ from the bottom to the top respectively. One can see that instead of the unbouned growth (as in the CV case) the complexity returns to its unperturbed value.}
 \label{fig:WdW1}
\end{figure}

Working in the probe limit, we approximate the boundary of the patch to be the pure $AdS$ null rays\footnote{An important thing to point out is that in the CA case we have to deal not with a constant time slice of the bulk, but with a patch of spacetime. As a result, the perturbative approximation might be not so perfectly accurate anymore. Still, at small $m$ it is good enough to capture the main qualitative features of the dynamics.}. We regularize the boundary by a small shift, $z_{bdy} = \varepsilon$, and consider null rays emanating from the boundary at time $t=\tau$: $z=\varepsilon+t-\tau$ and $z=\varepsilon-t-\tau$. We sketch the WDW patch in Fig.\ref{fig:WdW1}.

As we take into account only the matter term in the action, the non-trivial time dependence in this approximation is related to the fact that at different times $\tau$ the part of particle trajectory intersecting the WDW patch is different. 
Action of massive particle \eqref{part} evaluated on the part of its worldline between time moments $t_1$ and $t_2$ ($t_1<t_2$ for definiteness) is
\bea\label{eq:CAint}
S=-m\int_{t_1}^{t_2} \frac{\alpha dt}{\alpha^2+t^2}.
\eea
It intersects the ``left'' boundary of the WDW patch (Fig. \ref{fig:WdW1}a) at $t_1=\frac{\tau^2-\alpha^2}{2\tau}$, and $t_2=\infty$. Thus, integrating r.h.s. of \eqref{eq:CAint} we obtain
\bea\label{eq:CAtot}
\Delta\cC_{p}=-h  \left(1+\frac{2}{\pi} \arctan\left(\frac{\alpha }{2 \tau }-\frac{\tau }{2
   \alpha }\right) \right),
\eea 
where we used the relation $m=2h$. Note that due to the negative sign in the definition of massive particle action \eqref{eq:CAint} right after the quench the action complexity gains a negative correction to its vacuum (pure $AdS$) value (a similar effect has been observed in \cite{Chapman:2016hwi}). Around $t \rightarrow 0$, the complexity grows linearly with a pace defined by the total energy of the perturbation
\be \label{CAtotapp}
\Delta\cC_{p}\approx -2   h+ \frac{4 h  }{\alpha \pi }\tau=-2 h + \frac{2E}{\pi}\tau,
\ee
while in the long-term it approaches the equilibrium value as
\be 
\Delta\cC_{p}\approx -\frac{4 \alpha  h }{\pi \tau }
\ee
The time dependence of complexity \label{CAtot} is shown in Fig.\ref{fig:WdW1}.b for different values of $\alpha$. 

It is worthwhile to mention that the rate of complexification according  to \eqref{eq:CAtot} is
\be 
\frac{d\Delta\cC}{d \tau}=\frac{1}{\pi}\frac{4 \alpha h }{  \alpha ^2+  \tau^2}.
\ee 
From this equation or from \eqref{CAtotapp} we clearly see that the holographic local quench process saturates the Lloyd computational bound \cite{Brown:2015lvg} precisely at the initial time moment.

The long-term behavior drastically differs from that computed within the CV approach. Contra to the ``complexity always increases'' principle, instead of unbounded growth, we see convergence of complexity to its unperturbed value. The negative leading order correction in \eqref{CAtotapp} is another feature that deserves attention. Within the framework of holography and quantum field theory, the complexity of a state is meant to be complexity of {\it{formation}}, i.e. how much effort is required to create a desired state from some given reference state. Thus, if the the pure $AdS$ vacuum is taken to be the reference state, complexity of formation of any perturbation must be non-negative. The negative correction we derived poses a question whether the CA conjecture always deals with the complexity of formation or with a different kind of complexity.

\subsection{Subsystem complexity}

An extension of the CA conjecture onto the case of a subsystem in mixed state has been suggested in \cite{Carmi:2016wjl}. In this case, the bulk action has to be evaluated not over the full Wheeler-DeWitt patch, but rather over its intersection with the entanglement wedge corresponding to the boundary subsystem. Again, for simplicity we take our subsystem to be a single interval, and, as before, the Hubeny-Rangamani-Takayanagi surface is approximated by the one of the underformed $AdS$ spacetime. The resulting diamond shaped region of the bulk is presented in Fig.\ref{fig:WdW1}.b.

\begin{figure}[h!]
\centering
a)\,\,\includegraphics[width=4.5cm]{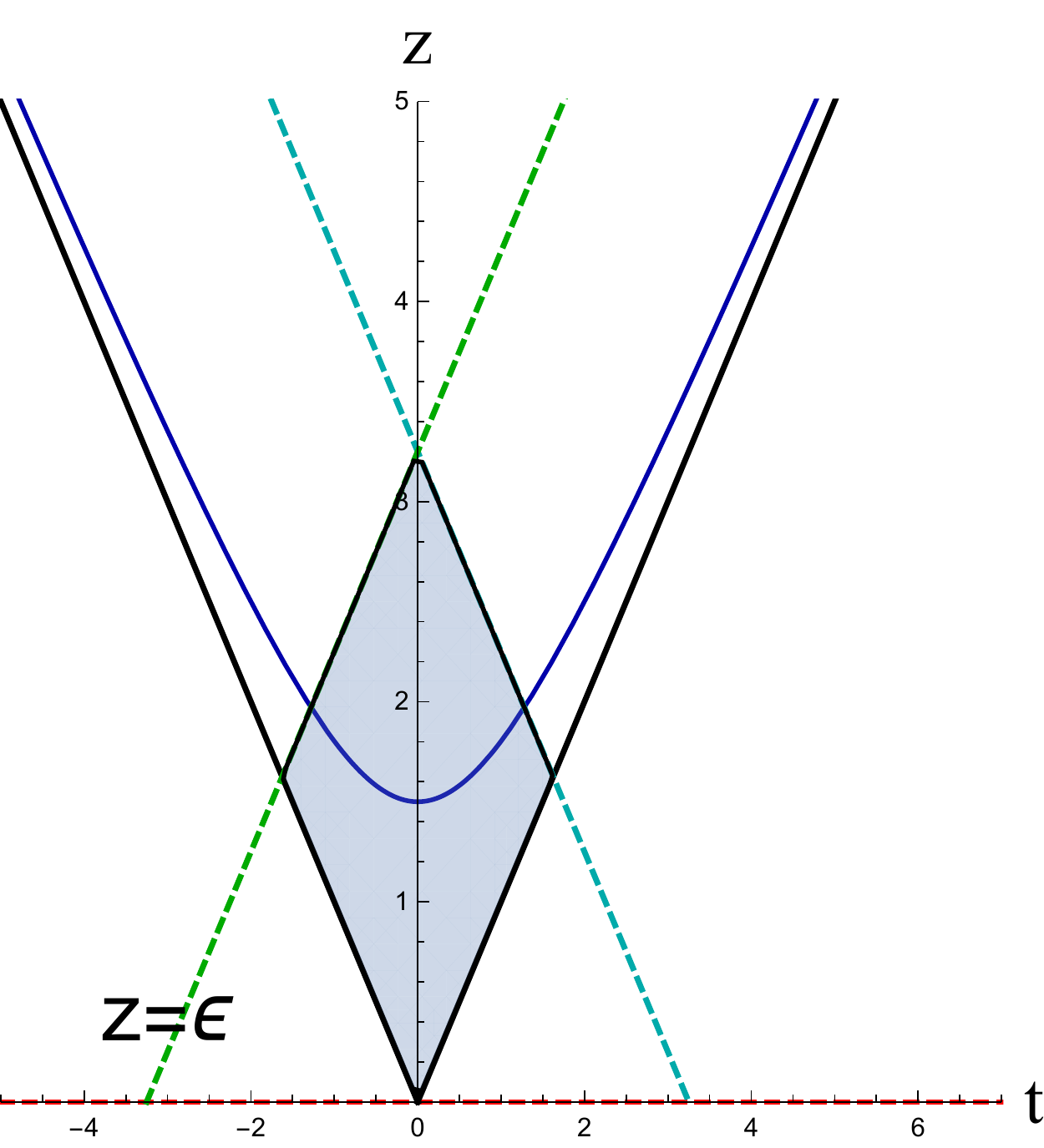}
b)\,\,\includegraphics[width=4.5cm]{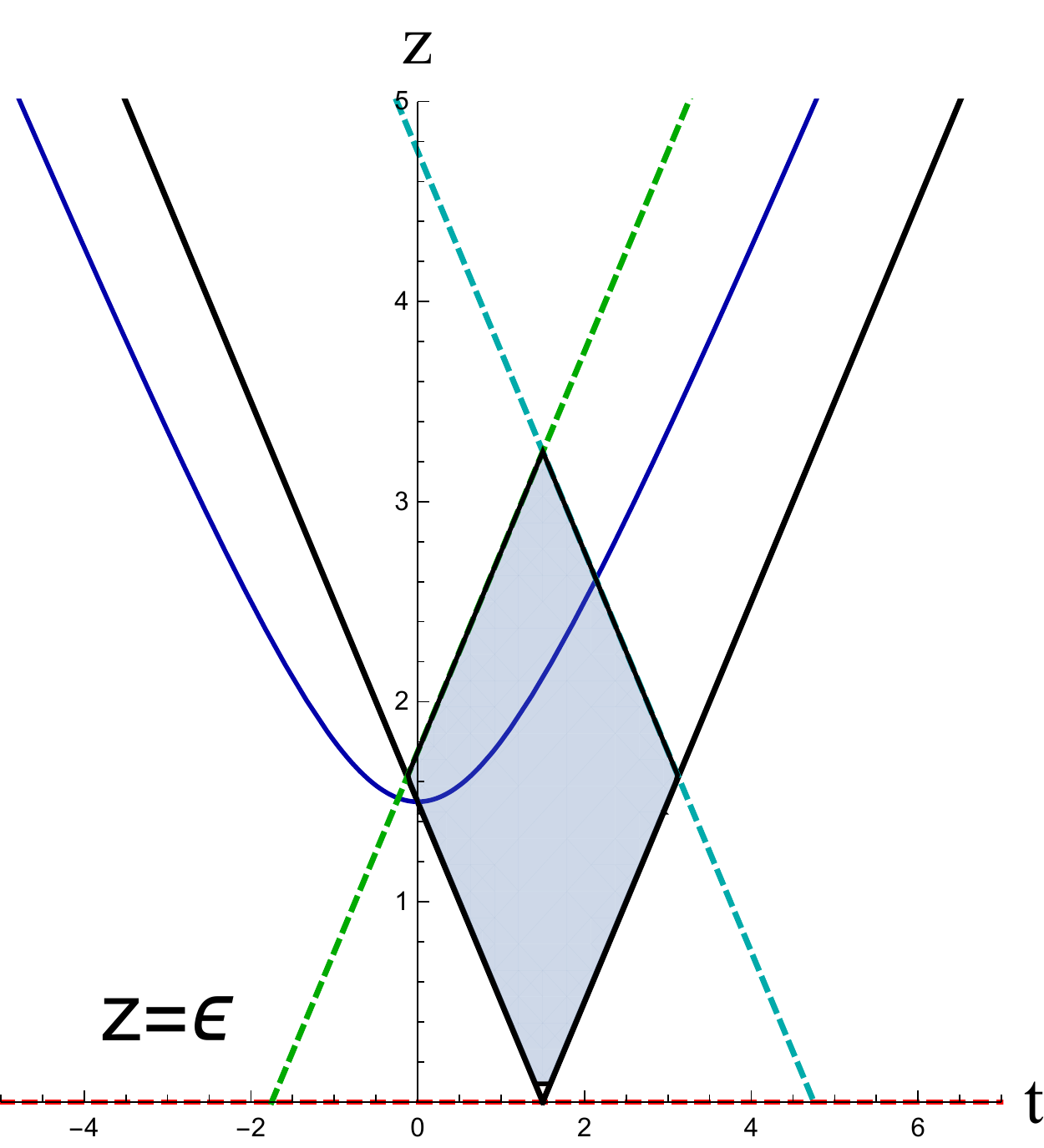}
c)\,\,\includegraphics[width=4.5cm]{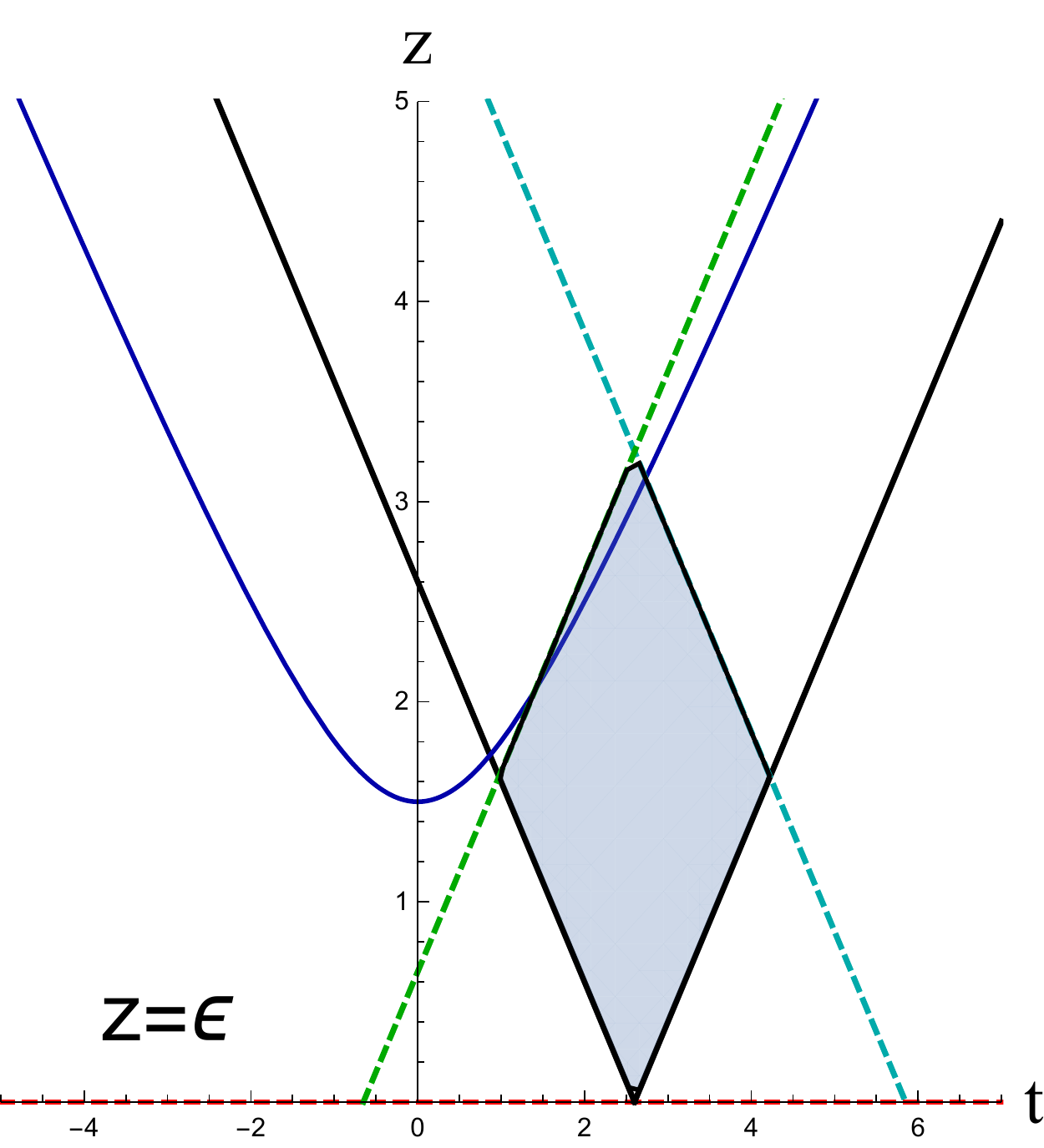}
 \caption{Possible different configurations of the WDW patch, the entanglement wedge, and the particle worldline (the blue curve). The solid region is the intersection of the WDW patch and the entanglement wedge. All plots show the case $\alpha<\ell/4$. The left plot corresponds to the early time evolution $\tau<\tau_{\text{crit}}$, the central plot is for $\tau_{\text{crit}}<\tau<\tau_{f}$, and the right one is for $\tau>\tau_f$.} 
 \label{fig:CA3}
\end{figure}

\begin{figure}[h!]
\centering
a)\,\,\includegraphics[width=5.5cm]{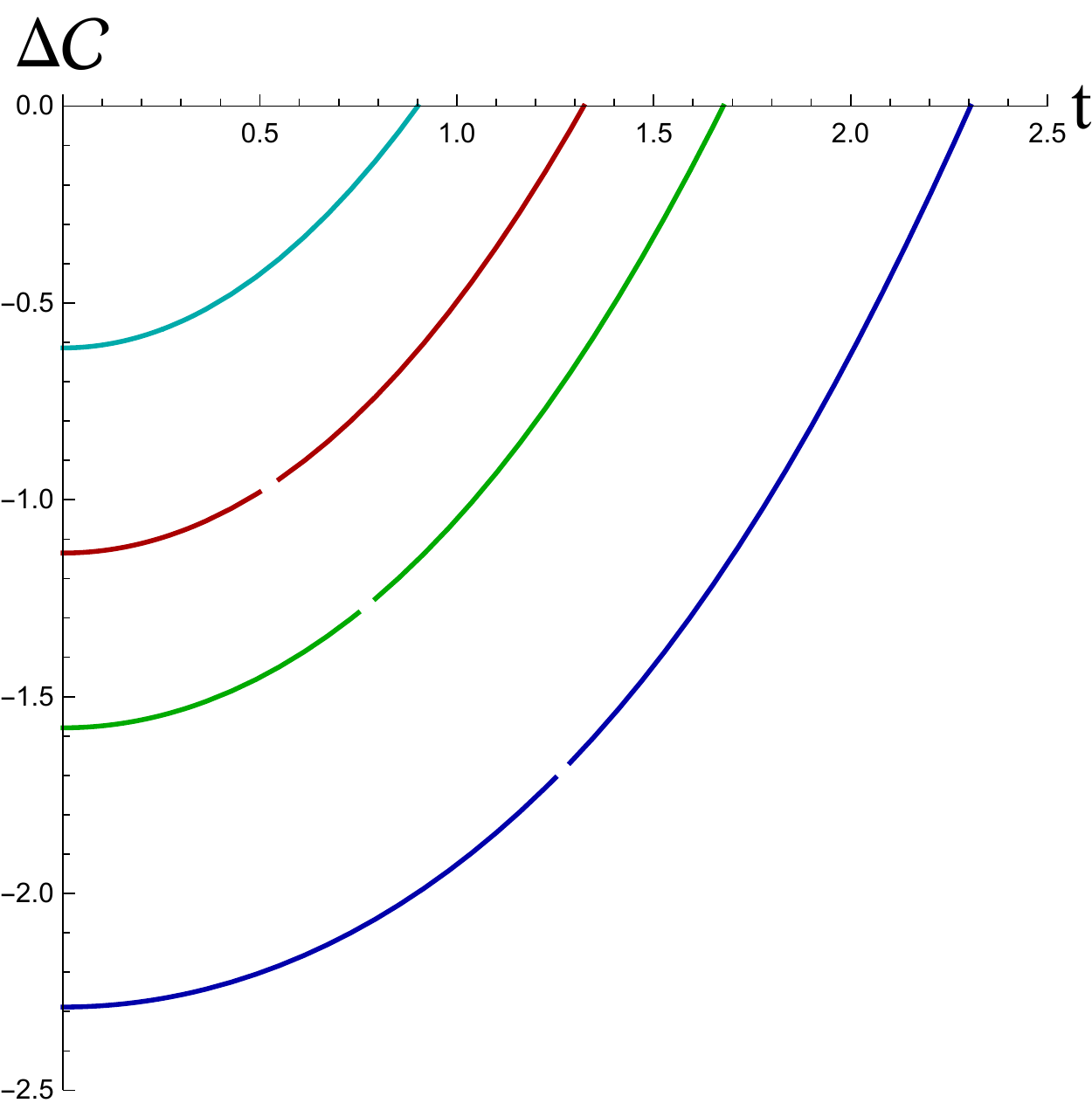}
b)\,\,\includegraphics[width=5.5cm]{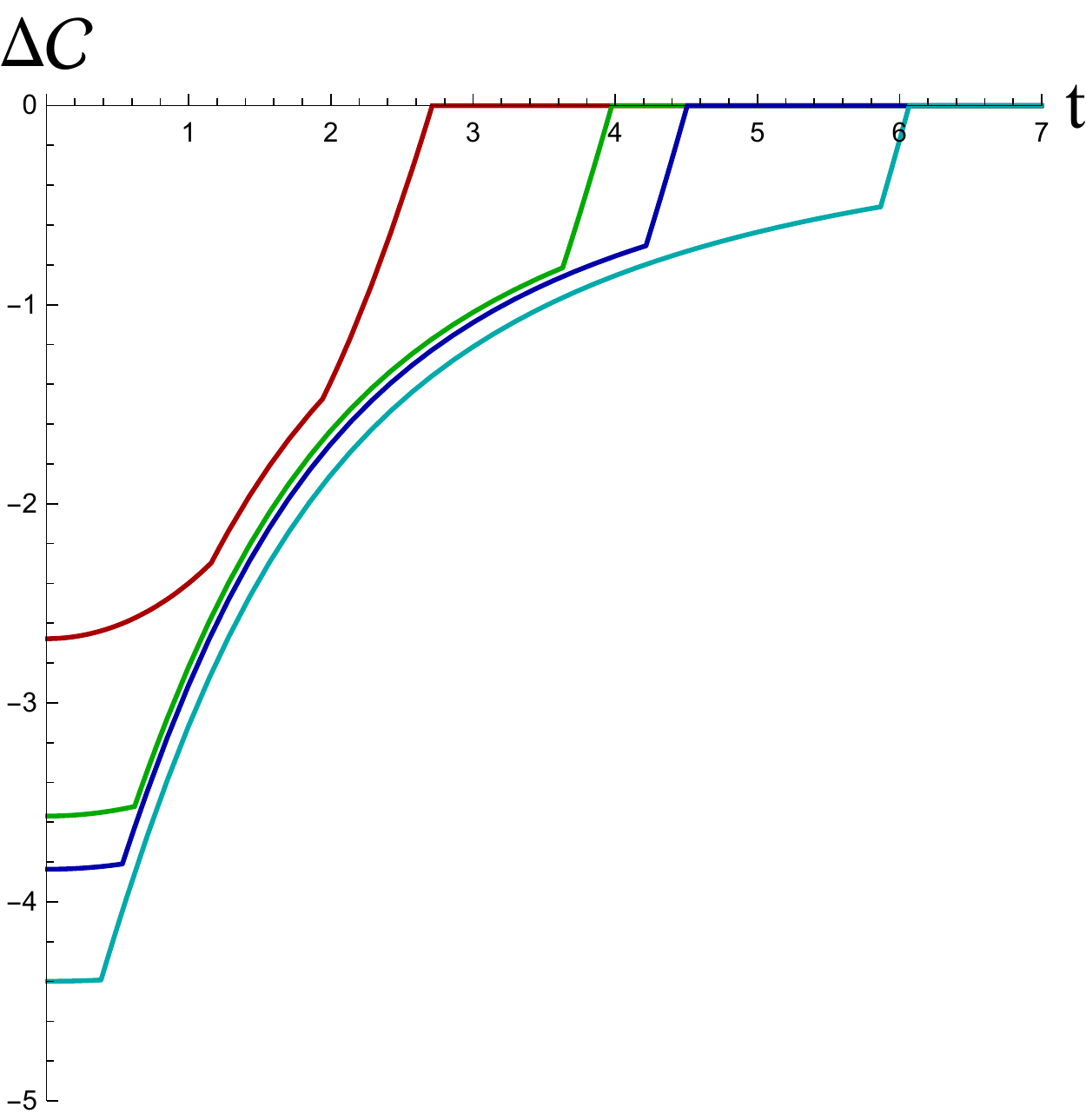}
 \caption{(a) Evolution of $\Delta \cC$ in the case of a mild quench, $\ell/4<\alpha<\ell/2$. Different curves correspond to $\ell=3.5,4,4.5,5.5$ (from top to bottom); $\alpha =1.5$. (b) The same quantity in the case of a sharp quench, $\alpha<\ell/4$. From top to bottom: $\ell=6.2, 8.5, 9.5, 12.5$; $\alpha=1.5$  }
 \label{fig:CAtl}
\end{figure}

Depending on the values of $\ell$ and $\alpha$, three different regimes are possible:
\begin{itemize}
\item If $\alpha>\ell/2$, the particle worldline does not intersect the diamond, and the complexity does not receive non-trivial corrections at all.
\item If $\ell/4<\alpha<\ell/2$, the particle worldline intersects the diamond across the boundaries formed by the entanglement wedge. We present the complexity time dependence for this case in Fig.\ref{fig:CAtl}.a.
\item If  $\alpha<\ell/4$, the structure of intersections between the particle worldline and the diamond is more complicated (see Fig.\ref{fig:CA3}). At early times $\tau<\tau_{crit}$, the intersection points again belong to the entanglement wedge boundaries (plot (a)). As the diamond evolves in time, after certain moment $\tau_{\text{crit}}$, the first intersection of the particle worldline with the diamond occurs at time $t_1$ at the boundary of the Wheeler-DeWitt patch, and leaves it at time $t_2$ from the entanglement wedge boundary (plot (b)). Finally, at the last stage, after some time $\tau_f$, the intersection picture changes once more to the initial one, - both intersection points belong to the boundaries of the entanglement wedge (plot (c)). The resulting evolution of complexity $\Delta \cC$ is presented in Fig.\ref{fig:CAtl}.b.
\end{itemize}
It is important to make a comment here. From the physical point of view, the quench should happen at $\tau=0$, and the fact that we take into account the part of the bulk particle worldline at negative times looks confusing. The necessity to continue the particle trajectory into the negative time domain is dictated by the condition of self-consistency of the local quench model, - then the holographic computation with metric \eqref{eq:long_metric} correctly reproduces the conformal field theory result \cite{Calabrese:2007mtj} at $\tau>0$. Before, when we considered the volume complexity (both of the total system and of the subsystem), and the action complexity of the total system, this model peculiarity did not matter. But here the negative time part of the trajectory directly affects the result. It is easy to remove this contribution by restricting the domain of integration in \eqref{eq:CAint} to $\tau\geq 0$ when computing the action complexity. The fact that the local quench is triggered by a heavy operator at $\tau=0$ does not go along with the time-reverse symmetric holographic picture, and the CA computation outlined above involves evaluation of the particle action at $\tau<0$, which is in some sense "unphysical". This time-reversal symmetry  is the drawback of the model. Nonetheless, this model and its generalizations are proven to reproduce different features of locally excited CFT \cite{Nozaki:2013wia,Asplund:2013zba,David:2017eno}. Thus we proceed in straightforward way including $\tau<0$ contribution, following the canonical CA proposal.
However, if we restrict the domain of integration to $\tau>0$, the qualitative behavior of complexity would not change much.

First, let us keep the negative time contribution to the action.
If the quench is not too sharp, $\ell/4<\alpha<\ell/2$, the region where the action is evaluated and the worldline are similar to the ones plotted in Fig.\ref{fig:CA3}.a, and
\bea\label{t1t2}
&&t_1=\frac{(\ell -2 \tau )^2-4 \alpha ^2}{4 ( 2 \tau-\ell )},\\
&&t_2=\frac{(2 \tau +\ell )^2-4 \alpha ^2}{4 (2 \tau +\ell )}.
\eea  
Calculating r.h.s. of \eqref{eq:CAint} for these times we obtain
\bea \label{CAl2}
&&\Delta\cC=-2 h \left(\text{arccot}\left(\frac{4 \alpha  (\ell -2 \tau )}{(\ell
   -2 \tau )^2-4 \alpha ^2}\right)+\text{arccot}\left(\frac{4 \alpha  (2
   \tau +\ell )}{(2 \tau +\ell )^2-4 \alpha ^2}\right)\right).
\eea
At early times, we find that the complexity grows quadratically
\be 
\Delta \cC=4 h  \arctan\left(\frac{\alpha }{\ell
   }-\frac{\ell }{4 \alpha }\right)+\frac{64 \alpha  h   \ell }{\left(4 \alpha ^2+\ell
   ^2\right)^2}\tau ^2.
\ee
In contrast with the volume complexity of an interval, here $\Delta \cC$ vanishes abruptly at time $\tau_{sat}=\frac{1}{2} \sqrt{\ell ^2-4 \alpha ^2}$.

If the quench is sharp, $\alpha<\ell/4$, the time dependence of complexity between $\tau=0$ and some $\tau=\tau_{\text{crit}}$ is the same as in the previous case. This is due to the fact that the structure of intersections between the particle worldline and the diamond remains the same (see Fig.\ref{fig:CA3}.a). This implies that the initial growth is also quadratic in time, but at the moment $\tau_{\text{crit}}$
\be \label{tcrit}
\tau_{\text{crit}}=\frac{1}{4} \left(\ell -\sqrt{\ell ^2-16 \alpha ^2}\right).
\ee
 the regime sharply changes. 
Then, after $\tau_{\text{crit}}$ and before certain time $\tau_f$, the ``entrance'' time $t_1=\frac{1}{2\tau}(\tau^2-\alpha^2)$, and the ``exit'' time $t_2$ is defined by \eqref{t1t2} (see Fig.\ref{fig:CA3}.b). So the time dependence of $\Delta \cC$ has the form
\be \label{CAell3}
\Delta\cC= 2h  \left(\arctan\left(\frac{\alpha ^2+\tau ^2}{2 \alpha  \tau
   }\right)-\arctan\left(\frac{(2 \tau +\ell )^2-4 \alpha ^2}{4 \alpha
    (2 \tau +\ell )}\right)\right).
 \ee
After time $\tau_f$ defined by
\bea 
\tau_f=\frac{1}{4} \left(\sqrt{\ell ^2-16 \alpha ^2}+\ell \right)
\eea 
the particle worldline intersects only the entanglement wedge boundaries (see Fig.\ref{fig:CA3}.c.). $\Delta \cC(t)$ is again described by \eqref{CAl2}, and vanishes at  $\tau_{sat}$. 

Thus it turns out that the time evolution of the action complexity exhibits non-smooth dependence on the relation between interval size $\ell$ and strength of the quench $\alpha$. This monotonous but discontinous (in derivative) behavior is in contrast with the much more "regular" behavior of the volume complexity, and we cannot exclude that it might be an artifact of the adopted approximation (that we took into account only the matter contribution to the bulk action, and ignored the possible issues related to the joints).

\section{The Lloyd bound}\label{sec:lloyd}
Inspired by different conjectures about the ultimate physical limit on the maximal speed of classical computation\footnote{I.e. on the speed of computing a desired state departing from a reference one by means of classical logical gates.} \cite{Anandan:1990fq}-\cite{lloyd_qg3}, the following bound on the speed of complexity growth (complexification) has been conjectured for the action complexity in  holography \cite{Brown:2015lvg}:
\be \label{eq:Lloyd}
{\cal R}<\frac{2E}{\pi},
\ee 
where ${\cal R}$ is the rate of complexification  ${\cal R}(\tau)=\frac{d\Delta \cC}{d\tau}$.

\begin{figure}[t!]
\centering
a)\,\,\includegraphics[width=6.5cm]{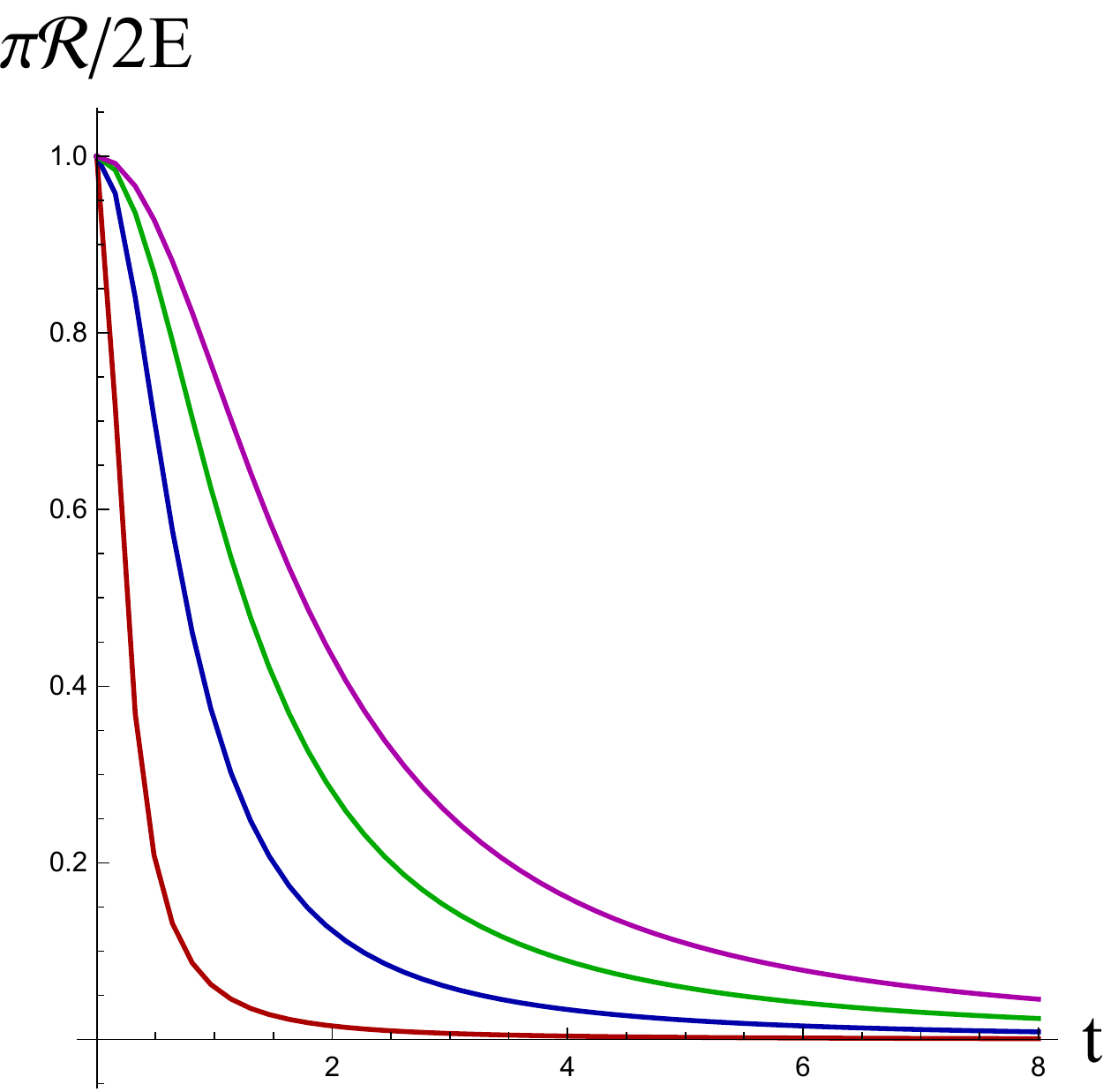}
b)\,\,\includegraphics[width=6.5cm]{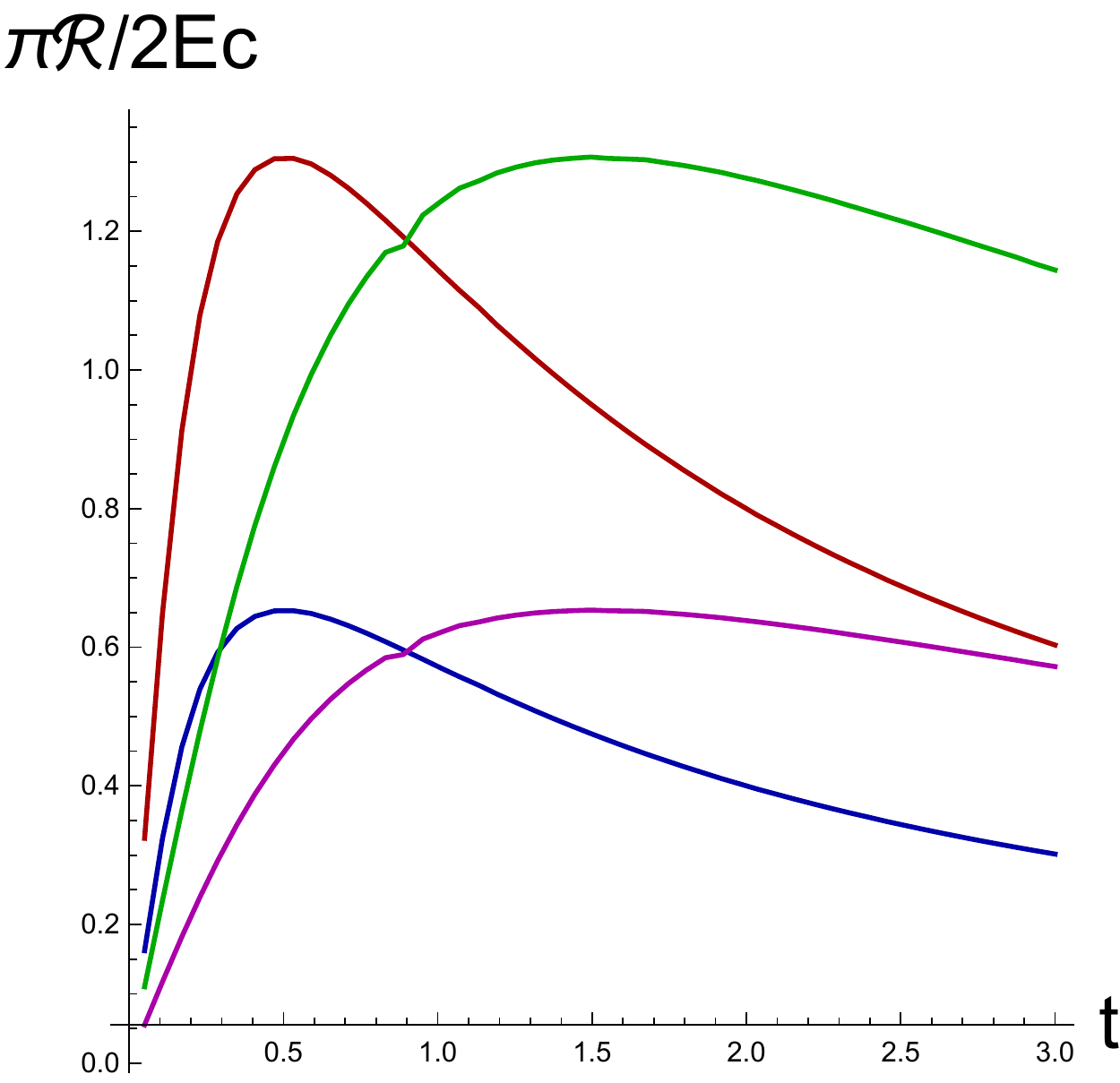}
 \caption{On the left plot, we present normalized complexification rate $\pi{\cal R}/2E$ calculated within the CA duality. Different curves correspond to different values of $\alpha=0.25,0.75,1.25,1.75$ from the left to the right. On the right plot, the ratio $\pi{\cal R}/(2Ec)$ calculated for the total system within the CV duality is shown, $c=24h/M$.  The red and and the blue curves have $\alpha=0.25$, while the green and the magenta ones - $\alpha=0.75$. $M=0.1$ for all curves. The top curves, violating the Lloyd bound, correpond to insertion of a lighter operator dimension, $h=0.1$, while the bottom ones - to heavier operator with $h=0.2$. } 
 \label{fig:R}
\end{figure}
If the CA conjecture is assumed, the complexification rate behaves in an unexpected and interesting way. As we already mentioned in the previous section, the complexification rate for the total system always saturates the Lloyd bound at the initial moment of time $\tau=0$ (independently of $\alpha$, $h$ and $c$), and then decreases. In Fig.\ref{fig:R}.a, we show this dependence for different $\alpha$. Recently the evolution of the action complexity following a global quench  has been considered in \cite{Moosa:2017yvt}. In contrast to the local quench results, in the global quench setting, the Lloyd bound is saturated soon after the local equilibrium in the system is established.

Within the CV conjecture, the complexification rate ${\cal R}$ evolves differently. To compute it, we need to fix the central charge, that has to be large since we deal with a holographic CFT, and the conformal dimension. Our approximation works well if $M<1$ (which means $c>24h$). We find that for fixed $h$ and $c$, and for different values of $\alpha$ (strength of the quench) the maximum of complexification rate ${\cal R}$ is reached at different times (later for larger $\alpha$), but has the same value. To discuss the Lloyd bound for the CV complexification rate, we need to take care. The bound \eqref{eq:Lloyd} has been conjectured for the action complexity, and should not be directly applied to the volume one as the latter is proportional to the central charge, which can be taken arbitrarily large. Therefore we suggest to normalize it by factor of $c$. We present the evolution of the ratio $\pi{\cal R}/(2Ec)$ in Fig.\ref{fig:R}.b. We can see that for larger conformal dimensions the ``normalized'' bound is satisfied, $\pi{\cal R}/(2Ec)<1$, while for smaller $h$ it is violated. That is a very intuitive result, since originally the Lloyd bound has been proposed within the context of classical computations, and the regime of large $h$ and $c$ corresponds precisely to the case when the holographic dual of a CFT is classical.

\section{Discussion}\label{sec:disc}
In this paper we have studied how the holographic complexity of a state in $1+1$-dimensional conformal field theory evolves after a sudden local perturbation, which can be seen as an insertion of heavy operator of dimension $h$. Following \cite{Nozaki:2013wia}, we take the holographic dual of this process to be the Poincare patch of $AdS_3$ perturbed by a massive point-like particle. Focusing on the case of weak perturbation, we analytically compute the evolution of holographic complexity of the total system and of an interval subsystem within both the CV and the CA proposals. 
As we emphasized in the introduction, we performed the analysis with the main aim to employ holography to gain some understanding of the emergent effective complexity as it is intuitively perceived. Now we would like to summarize the results and make an attempt to conceptualize the observed dependences and phenomena by putting them into a broader context of the general problem of complexity in science.

Our results for the volume complexity of the total system can be summarized as follows - once quenched, the system exhibits steady growth of complexity, quadratic in time in the beginning, and slower at large times. This is similar to what has been found before in the case of global quench \cite{Moosa:2017yvt}. Comparing the complexity evolution with the evolution of entanglement entropy of semi-infinite bipartition of the system, we find that at the initial stage the complexity is proportional to the entanglement, while at later times this dependence is nonlinear (see Fig.\ref{fig:cwhole}). 
    
Time-dependent volume complexity of an open subsystem is more interesting, and has certain features that are important to highlight. Initially it increases in time, but after passing the maximum smoothly but quickly returns to its unperturbed value\footnote{A similar behavior has been found in the case of global quench \cite{Moosa:2017yvt}. While global thermal quench does not create a non-trivial regular pattern, it is a very natural that complexity maximizes when the system has thermalized in the ultraviolet, while retaining information about the original vacuum state at the macroscopic scales. This kind of interscale diversity is very typical for systems of high effective complexity \cite{dissimilarity}}. Over the course of the evolution, we observe a non-trivial relation between the volume complexity of an interval and its entanglement entropy and integrated absolute entanglement density $\mathcal N$, which we use as a possible quantitative measure of randomness of a state. There is a regime when complexity of the interval has already reached its maximum and decreases, while the entanglement entropy and $\mathcal N$ keeps growing. This kind of behavior is in line with the notion of effective complexity that should be maximal not for a random structure, but rather for a highly-organized one. That leaves a room for possible future connections between holography and the theory of physical (non-computational) complexity \cite{Lloyd1}-\cite{Lloyd2}.

The CA conjecture analysis shows a somewhat different behavior. First of all, because of the minus sign in the canonical definition of a massive particle action, the quench causes a negative correction to the equilibrium value of complexity, which contradicts the basic intuition that complexity of formation of a new state from a given reference state should be non-negative. Then, for the total system, the action complexity grows linearly at the early times (in contrast with the quadratic growth of the volume complexity), and asymptotically approaches its original value as $t\rightarrow \infty$. For a finite interval, evolution of the action complexity depends substantially on its size. If the interval is too small as compared to $\alpha$ ($\ell<2\alpha$), its complexity doesn't change at all upon the quench. For intervals of intermediate sizes, the correction to action complexity demonstrates quadratic dependence on time, and vanishes abruptly at saturation time. Finally, if the interval is large enough ($\ell>4\alpha$), $\Delta {\cal C}(t)$ sharply changes the regime twice over the course of evolution before approaching zero. It is quite likely that this irregular behavior is an artifact of the adopted model of holographic local quench, and resolving the point-like locality of the perturbation will lead to a smoother evolution. Still, it is worthy to point out that the CV conjecture is more tolerant to this type of model shortcomings.

Another thing interesting to pay attention to is the time dependence of the complexification rate ${\mathcal R} = d \Delta {\mathcal C}(t)/dt$ for the total system, which turns out to be very different for the volume and the action complexities. Here, the CA conjecture leads to a result we find important to comment on. Independently of parameters of the quench (conformal dimension $h$, ``sharpness'' $\alpha$), at the very first moment the Lloyd bound is exactly saturated, and then monotonously decreases. To find an intuitive interpretation of  this behavior, let us look at it from the perspective of the theory of self-organized criticality \cite{Bak}. This theory is one of the main paradigms of complexity formation in natural systems, and relates the natural (geological, biological etc.) patterns to emergent scale invariance. While this view seems to be oversimplified, as a truly ``complex'' system should demonstrate non-trivial hierarchy of levels rather than to self-reproduce itself at different scales, 
it is nice to see how our observation fits into this picture complementing it. 
Although here we deal with a quantum conformal theory and not with a classical scale-invariant model, sticking to the same general philosophy we can say that criticality itself does not imply complexity, but serves as {\it a seed} of it: when hit with a local perturbation, a scale-invariant system demonstrates the maximal possible rate of complexity growth. 

The CV complexification rate for the total system changes non-monotonously in time, and violates the normalized Lloyd bound for small conformal dimensions $h$ of the quenching operator. For large $h$ this violation is absent. This is consistent with the picture that the classical bulk description is valid when both the large central charge $c$ and the operator dimension $h$ are large enough, and small $h$ requires computing quantum corrections in the bulk. Also an interesting fact is that the maximal complexification rate depends only on $h$ and not on $\alpha$, while the latter sets the time when this maximum is reached.

To sum up the discussion, we would like to say that, while we should not expect a unique notion of complexity to exist, apparently holography is indeed capable of capturing certain traits of effective physical complexity, and the CV and the CA definitions reflect different but equally relevant aspects of this concept. To further proceed on this way, one can pursue different directions. One thing we are willing to understand better is complexity of more non-trivial holographic patterns than just a pair of excitations. In \cite{DeJonckheere:2018pbi}, an approach to study arbitrary inhomogeneous quenches in $AdS_3/CFT_2$ has been suggested, which seems to be a convenient tool for the analysis of complexity evolution of different structures. Another alluring connection between holography and the theory of physical complexity could be established on the basis of the complex network representation of a quantum state proposed in \cite{CarrNetwork}. The idea is to define complexity of a wave function in terms of characteristics of the corresponding classical network, which nodes are associated with the degrees of freedom (sites of a spin chain, for example), and links are weighted with the values of inter-site mutual information (in the continuum case, such a network can be defined via discretization). Relating the network and the holographic complexities of the same state, one can shed a brighter light on the idea of holographic complexity. We hope to address some of these issues in the future.

\appendix

\section{ Holographic dual of local quench}\label{appA}
In this appendix, we derive the explicit form of the metric dual to the local holographic quench.
The global $AdS_3$ metric deformed by a static point particle of mass $m$ is given by
\be \label{appmetr1}
ds^2=-d\tau^2
   \left(L^2-M+R^2\right)+R^2 d\phi^2+\frac{L^2 dR^2 }{L^2-M+R^2}
\ee
where $M=8mGL^2$. For simplicity we take $L=1$.

Then, we obtain the holographic dual of the local quench by applying to \eqref{appmetr1} the following coordinate transformation
\bea\label{appmap}
&&\phi =\arctan\left(\frac{2 \alpha  x}{\alpha ^2+t^2-x^2-z^2}\right)\\
&&\tau =\arctan\left(\frac{2 \alpha  t}{\alpha ^2-t^2+x^2+z^2}\right)\\
&&R=\frac{\sqrt{\alpha ^4+2 \alpha ^2 \left(t^2+x^2-z^2\right)+\left(-t^2+x^2+z^2\right)^2}}{2 \alpha  z}.
\eea
The resulting metric is
\bea \label{eq:long_metric}
&&ds^2=\frac{1}{z^2}\frac{\left(\alpha
   ^2 dx-2  t x dt+dx \left(u-z^2\right)+2 x z dz
   \right)^2}{\alpha ^4+2 \alpha ^2
   \left(u-z^2\right)+\left(z^2-v\right)^2}-\\\nn
&&-\frac{1}{z^2}\frac{\left(\alpha ^4+2
   \alpha ^2 \left(u+z^2(1-2 M) \right)+\left(z^2-v\right)^2\right)
   \left(\alpha ^2 dt+ \left(u+z^2\right)dt-2 t (xdx
   +zdz )\right)^2}{\left(\alpha ^4+2 \alpha ^2
   \left(u+z^2\right)+\left(z^2-v\right)^2\right)^2}\\\nn
&&\frac{1}{z^2}\frac{\left(\alpha ^4 dz+2 \alpha ^2 (udz -z(t dt +x dx ))+\left(v-z^2\right) \left(-2 t z dt +2 x z dx +\left(v+z^2\right)dz
   \right)\right)^2}{\left(\alpha ^4+2 \alpha ^2
   \left(u-z^2\right)+\left(z^2-v\right)^2\right) \left(\alpha ^4+2
   \alpha ^2 \left(-2 M
   z^2+u+z^2\right)+\left(z^2-v\right)^2\right)},
\eea
where we introduced $u=t^2-x^2$ and $v=t^2+x^2$. One can check that for $M=0$ \eqref{eq:long_metric} is always the metric of the Poincare patch, independently of $\alpha$.
\section{Perturbative computation of  holographic entanglement entropy}\label{pert}

In this section, we outline the perturbative calculation of the holographic entanglement entropy $S$ of a single interval in the $AdS_3$ spacetime deformed by a massive point particle. To derive the exact expression for $S$ requires solving the equations of motion of the Hubeny-Rangamani-Takayanagi surface, which is doable but somewhat cumbersome. Instead, we consider $M$ to be a small parameter ($M\ll L^2$), and approximate the HRT surface by that of the empty $AdS$ spacetime:
\bea
z_{HEE}=\sqrt{\ell^2-x^2}.
\eea
The induced metric on this curve inherited from background \eqref{eq:long_metric} is
\bea
&&ds_{HEE}^2=g_{HEE}dx^2,\\
&&g_{HEE}=\frac{ \left(\left(\alpha ^2+t^2-\ell ^2\right)^2-\frac{x^2 \left(\alpha ^4+2 \alpha ^2 \left(t^2+\ell ^2\right)+\left(t^2-\ell
   ^2\right)^2\right)^2}{\left(x^2-\ell ^2\right) \left(\alpha ^4+2 \alpha ^2 \left(2 M x^2-2 M \ell ^2+t^2+\ell ^2\right)+\left(t^2-\ell
   ^2\right)^2\right)}\right)}{\left(\ell ^2-x^2\right) \left(\alpha ^4+2 \alpha ^2 \left(t^2+2 x^2-\ell ^2\right)+\left(t^2-\ell
   ^2\right)^2\right)}.
\eea
The resulting deviation of the entanglement entropy from the $AdS$ vacuum value is
\bea
\Delta S(\ell,t)=\int_0^{\ell}\Big(\sqrt{g_{HEE}}-\frac{ \ell}{\ell^2-x^2}\Big)dx.
\eea
Up to the first order in $M$:
\bea
\Delta S(\ell,t)\approx \frac{M}{4\alpha\ell} \left(2 \alpha  \ell -\left(\alpha ^2+t^2-\ell ^2\right) \arctan\left(\frac{2 \alpha  \ell }{\alpha ^2+t^2-\ell ^2}\right)\right).
\eea

\section{Initial growth}\label{sec:ingrowth}
To derive the $t \rightarrow 0$ asymptotic of interval volume complexity $\Delta {\cal C}$
\bea
\Delta {\cal C}(\ell,t) \approx {\cal C}_0+{\cal C}_1 t^2,
\eea
we take the constant time volume $\Sigma$ of metric \eqref{eq:long_metric} (as defined in \eqref{DV}), and expand it in $M$ and $t$:
\bea\label{exp1}\nn
&&\Sigma \approx \frac{2 \alpha ^2 r}{\left(\alpha ^2+r^2\right)^2}-\frac{4 \alpha ^2 M r \left(\alpha ^6-4 \alpha ^4 r^2+3 \alpha ^2
   r^4+r^2 \left(3 \alpha ^4-8 \alpha ^2 r^2+r^4\right) \cos (2 \phi )-4
   r^6\right)}{\left(\alpha ^2+r^2\right)^4 \left(\alpha ^4+2 \alpha ^2
   r^2 \cos (2 \phi )+r^4\right)}t^2
\eea
where we used radial parametrization $z=r \sin \phi$ and $x=r \cos \phi$.

Integrating the first term in this expression over the circle of radius $\ell$, we get the constant in time term
\bea
{\cal C}_0 =M\frac{\pi  \ell ^2}{\alpha ^2+\ell ^2},
\eea
In turn, the next order coefficient ${\cal C}_1$ is 
\bea
&&{\cal C}_1=\left\{\begin{array}{c}\frac{2 \pi  M  \ell ^2 (\ell^2-\alpha^2  ) }{\left(\alpha ^2+\ell ^2\right)^3},\,\,\,\,\,\,\alpha>\ell \\ \frac{\pi  M  \left(\alpha ^6-5 \alpha ^4 \ell ^2+3 \alpha ^2 \ell
   ^4+\ell ^6\right)}{2 \alpha ^2 \left(\alpha ^2+\ell ^2\right)^3}, \,\,\,\,\,\,\alpha<\ell. \end{array} \right.
\eea

\section{Complexity-action  conjecture}
In this appendix we give a brief review  of the details of the CA conjecture formulation. The CA formulation for a total system and for subregion (single interval) is given. 
\subsection{A total system}
First, consider complexity of the total system defined as a fixed time slice of the boundary at some $t=t_0$. The action then has the form
\be \label{appS1}
S=\frac{1}{16 \pi G} \int \sqrt{-g}({\cal R}-2\Lambda)dx^{d+1} +\frac{1}{16 \pi G}S_{matter}, \ee
where $S_{matter}$ is the matter action.
Complexity $\cC$ is identified  with action \eqref{appS1} evaluated over the special wedge called Wheeler-DeWitt patch. This patch is defined as a union of all spatial curves anchored on the boundary at $t_0$, or, equivalently, as a bulk region ${\cal M}$ bounded by light rays emanating from the boundary at $t_0$. 

\begin{figure}[t!]
\centering
\,\,\includegraphics[width=6.5cm]{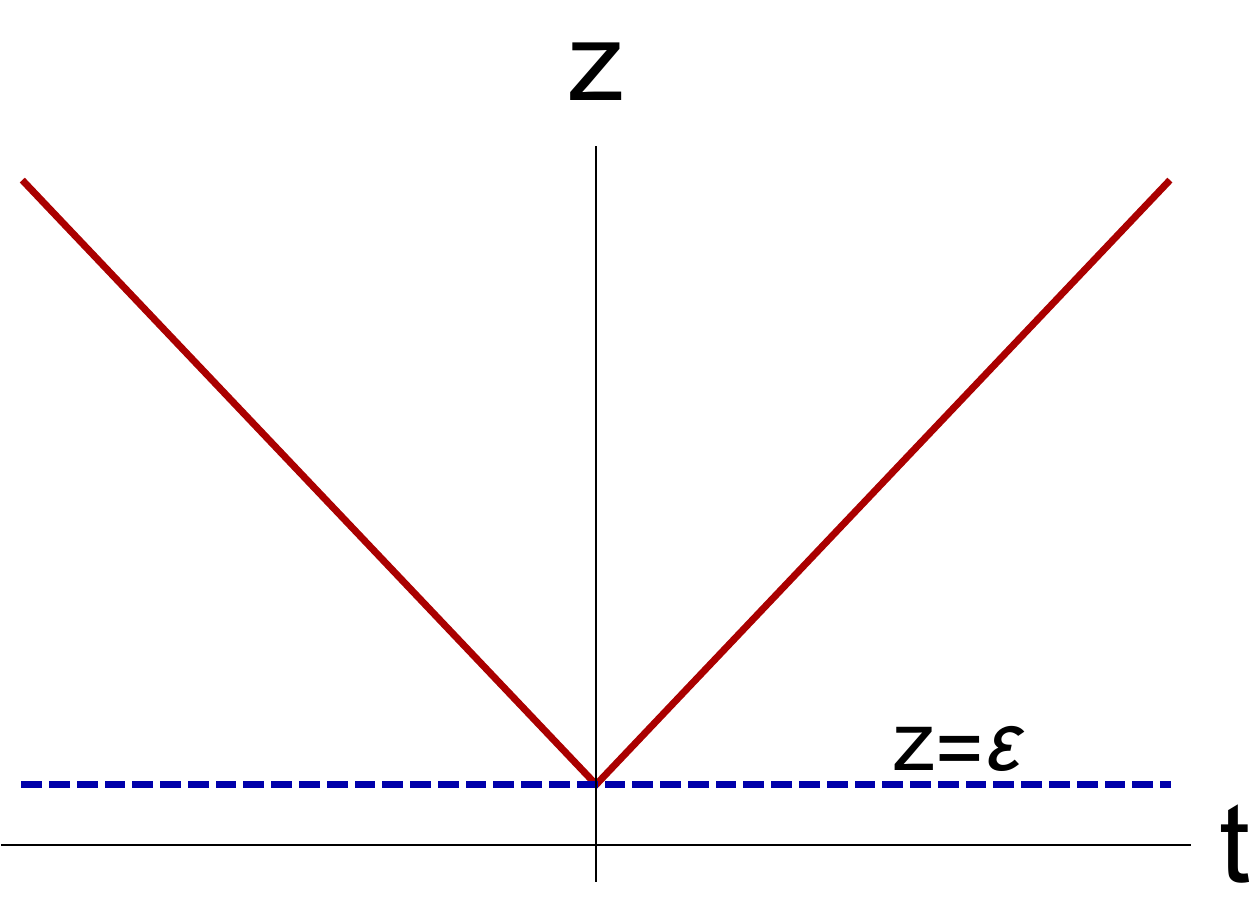}
\,\,\includegraphics[width=6.5cm]{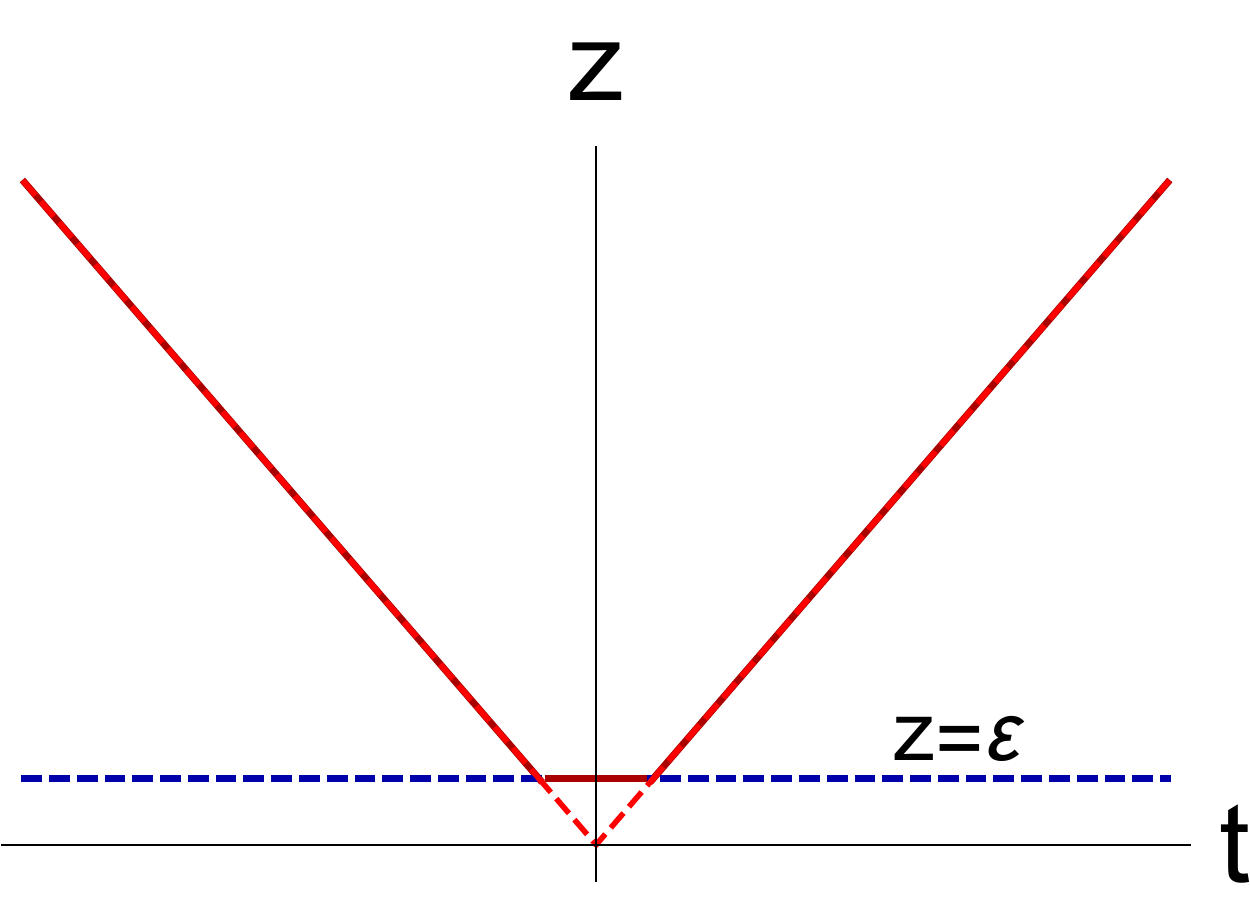}
 \caption{ The different WDW patch regularizations.} 
 \label{fig:appWdWt}
\end{figure}

When restricted to this region, action \eqref{appS1} should be modified to include so called joint, surface and boundary terms (their meaning will be explained below). The action is divergent, and two different regularization schemes can be employed. One approach is to place the edge of this patch on the $z=\varepsilon$ regulator (see the left plot of Fig.\ref{fig:appWdWt}).  The  second prescription is to place it right on the boundary $z=0$, but to introduce a cutoff at $z=\varepsilon$ (right plot of Fig.\ref{fig:appWdWt}). For simplicity, we call the first prescription ``lift regularization'', and the second one - ``cutoff regularization''. In the limit $\varepsilon \rightarrow 0$, we obtain the same result.  However, at finite $\varepsilon$  some terms may differ. For example, in the cutoff regularization scheme, the Gibbons-Hawking term is present, while in the lift regularization, it is absent. Also the joint terms are different. In this paper we use the lift regularization.

This said, we obtain:
\bea \label{appS2}
\nn
S_{WdW}&=&\frac{1}{16 \pi G} \int_{{\cal M}} \sqrt{-g}({\cal R}-2\Lambda)dx^{d+1} +\frac{1}{16 \pi G}\int_{{\cal M}}{\cal L}_{matter}dx^{d+1}+\frac{1}{8 \pi G}\int_{\cal S} \sqrt{\gamma}\log |\frac{k \bar k}{2}|dS+\\
&+&\frac{1}{8\pi G}\int\sqrt{\gamma}\Theta\log |L \Theta| dS d\lambda,\\
\Theta&=&\partial_{\lambda}\log \sqrt{\gamma}.
\eea
The first two term are the on-shell action of the gravity and the matter contribution from \eqref{appS1}. The third term is the joint term, and the fourth term is included to restore the reparametrization invariance of this action. The joint term corresponds to the intersection of two null-surfaces, and integration over $\cal S$ means the integration over the  induced metric on the joint. Here $\lambda$ is the affine parameter on the null boundaries of the WDW patch, $\gamma$ is the induced metric on constant $\lambda$ cross-section, and $k$ and $\bar k$ are future-directed tangents to the null surfaces. 
\subsection{Gravitational action in the Wheeler-DeWitt patch global AdS deformed by a static particle}\label{sec:WdW}
To elaborate on the notion of action complexity, let us study a simple example. Consider metric of the following form
\bea
ds^2&&=-f(R)dt^2+\frac{L^2dR^2}{f(R)}+R^2d\phi^2,\\
f(R)&&=A^2+R^2,
\eea
where $A=\sqrt{L^2-M}$. When $0<M<1$, this is the metric of global $AdS_3$ spacetime deformed by a static point particle. 
The CA prescription relates the complexity of a state in the boundary field theory to the bulk action evaluated in the Wheeler-DeWitt (WdW) patch bounded by two light-like surfaces
\be 
t={{\cal F}}_{\pm}=\frac{\pi}{2}\frac{L}{A}\pm \frac{L}{A}\arctan(\frac{R}{A}).
\ee 
The action to be evaluated comprises four terms:
\bea
&&S_{WDW}=S_{GR}+S_{joint}+S_{rep}+S_{particle}.
\eea
The  gravitational term is 
\begin{gather}\nn
S_{GR}=2\cdot\frac{1}{16\pi G}\int \sqrt{-g}({\cal R}+2)dzd\phi dt=2\frac{-4 \cdot 2\pi}{16\pi G L^2}\int_{0}^{r_m}  r dr\int_{0}^{{{\cal F}}_{-}}dt\\
\approx 2\frac{4 \cdot 2\pi}{16\pi G}\left(\frac{\pi  A}{4}-r_m\right)=\frac{\pi  A}{4 G}-\frac{r_m}{G},\,\,\text{as}\,\,r_m\rightarrow\infty
\end{gather}
where ${\cal R}=-6/L^2$, $\sqrt{-g}=L R$ and $r_m$ is the near-boundary cut-off, and we use the time-reversal symmetry to perform the integration.

The null joint contribution is
\bea
S_{joint}=\frac{1}{8\pi G}\int\sqrt{\gamma}\log{\frac{|k \cdot \bar k|}{2}}d \phi\Big|_{r_m} \label{eq:Snj}
\eea
where $k$ and $\bar k$ are vectors tangent to the null surfaces, and
\bea
&&\frac{1}{2}|k \bar k |= \frac{1}{2} g^{\mu \nu} k_{\mu} k_{\nu}=L\frac{\alpha \beta}{f},
\eea
where $\alpha$ and $\beta$ are arbitrary constants arising due to the fact that there is no canonical way to normalize a null vector \cite{Reynolds:2016rvl}. Action term \eqref{eq:Snj} takes the form
\bea\nn
&&S_{joint}=\frac{2\pi}{8\pi G} r \log L\frac{\alpha \beta}{f}\Big|_{r_m}=-\frac{2\pi}{8\pi G}r_m\log((1+r_m^2)\frac{1}{\alpha\beta})\approx\\
&&\approx-\frac{2\pi}{8\pi G}r_m\log(\alpha\beta L)+2r_m\frac{2\pi}{8\pi G}\log r_m
\eea

Finally, to compute the term required by reparametrization invariance we need to introduce affine parameters $\lambda,\,\, \bar\lambda$ on the null boundaries of the WDW patch:
\bea
&&\lambda=-\frac{r}{\alpha},\\
&&\bar\lambda=\frac{r}{\beta}. \nn
\eea
Then
\be
S_{rep}=S_{\Theta_{\alpha}}+S_{\Theta_{\beta}},
\ee
where
\begin{gather}
\Theta_{\alpha}=\partial_{\lambda}\log\sqrt{\gamma}=-\frac{\alpha}{r}, \\
S_{\Theta_{\alpha}}=\frac{2\pi}{8\pi G}\int \sqrt{\gamma}\cdot \Theta \log |\ell \Theta| d\lambda= \nn \\
=\frac{2\pi}{8\pi G}\int_0^{r_m} \log\frac{\alpha L}{r}dr d\phi=\frac{2\pi }{8\pi G}\left(r_m \left(\log \left(\frac{\alpha L}{r_m}\right)+1\right) \right),\nn
\end{gather}
and $\Theta_\beta$ is identical to $\Theta_\alpha$ up to changing $\alpha \rightarrow \beta$.

Summing up all the terms we obtain the answer
\be\label{eq:WdWstat}
S_{WDW}=\frac{1}{G}\frac{\pi \sqrt{L^2-M}}{4}-\frac{1}{2}\frac{1}{G}r_m + S_{particle}.
\ee
Note, that the ambiguity concerning $\alpha$ and $\beta$, and logarithmic divergent terms are cancelled in the total sum.

\subsection{Subregion complexity. A primer: Poincare AdS}
\begin{figure}[t!]
\centering
\,\,\includegraphics[width=7.5cm]{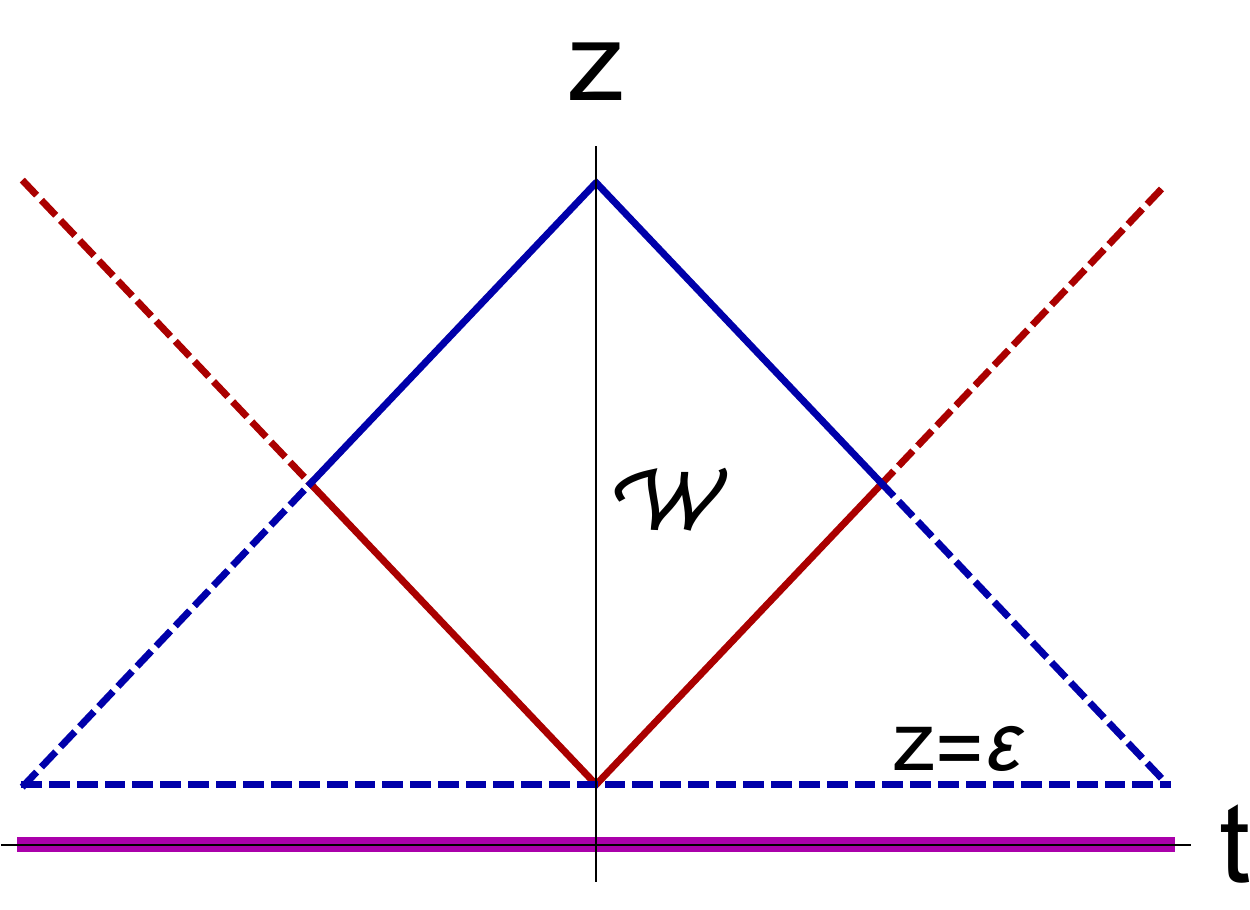}
 \caption{ The region $\cal W$ where one has to evaluate the gravitational action to compute the subregion complexity. The blue lines correspond to the entanglement wedge boundary, and the red lines are the WdW region null boundaries. Magenta line depicts the subregion of interest.} 
 \label{fig:appWdWent}
\end{figure}

To define the action complexity for subregion $A$, the bulk patch where the on-shell action is evaluated should be modified. It will be defined below, and the additional joints in principle should appear in the on-shell region. The complete analysis of these terms similar to what has been done for the action complexity of a full system \cite{Lehner:2016vdi} is still lacking, so we will restrict ourselves to studying the contributions due to volume terms in the action. In this appendix we follow \cite{Ben-Ami:2016qex}.

The modified WdW region is defined as an intersection of the WdW patch corresponding to the fixed boundary time moment $t_0$ and the region called the entanglement wedge associated with the subregion. The latter is defined as follows.
First, one has to define the Hubeny-Rangamani-Takayanagi surface $\chi_A$ associated to  $A$. Then, the entanglement wedge ${\cal E}_A$ is the causal domain of dependence\footnote{The causal domain of dependence $D[\Gamma]$ of some region $\Gamma$ is defined as the set of all points $P$ such that timelike curves passing through $P$ intersect $A$.} of the bulk subregion below the HRT surface $\chi_A$. The entanglement wedge is believed to encode the information about reduced density matrix of the subregion.

As an example, we will briefly outline the computation of the complexity of the radius $\ell$ sphere on the boundary of $AdS_{d+1}$. We follow the calculations from \cite{Ben-Ami:2016qex}. The $AdS_{d+1}$ metric is
\be 
ds^2=\frac{L^2}{z^2}\left(-dt^2+dz^2+d\bar x_{i}^2 \right).
\ee 
The HRT surface is
\be 
z^2+r^2=\ell^2,\,\,\,\,\,\,r=\sum_{i}\sqrt{x_i^2}.
\ee
Null boundary surface of the entanglement wedge is 
\be 
r^2+z^2=\ell-t,\,\,\,\,\,\,t>0.
\ee
The wedge where we have to compute the action is shown in Fig.\ref{fig:appWdWent}.

The gravitational action $S_{GR}$ has the form
\be 
S_{GR}=\left({\cal R}-\frac{d(d-1)}{L^2}\right)V,
\ee 
where the $V$ is the volume integral  expressed as
\be 
V=\frac{2\omega_{d-2}L^{d+1}}{d-1}\int_{0}^{\ell}\int_0^{\ell-t}\frac{((\ell-t)^{2}-z^2)^{(d-1)/2}}{z^{d+1}}dz=\frac{2\omega_{d-2}L^{d+1}}{d-1}\Big(\frac{1}{d^2}\frac{\ell^d}{\varepsilon^d}-\frac{d-1}{2(d-2)^2}\frac{\ell^{d-2}}{\varepsilon^{d-2}}+...\Big)
\ee
with $\varepsilon$ in the limit $\varepsilon \rightarrow 0$.

\section*{Acknowledgements}
We are thankful to Victor Kleptsyn, Alexander Krikun, Koenraad Schalm and Stanislav Smirnov for fruitful discussions on the related topics, and to Lincoln Carr for bringing to our attention the concepts of the theory of complex networks. This work is supported by the Russian Science Foundation (project 17-71-20154)

\end{document}